%% file: main.tex
\documentclass[letterpaper,twocolumn,10pt]{article}
\usepackage{usenix2019_v3}
\pdfoutput=1
\input{packages}

\input{preamble}

\makeatletter
\let\@fnsymbol\@arabic
\makeatother
\begin{document}
\hypersetup{
  linkcolor=purple,
  citecolor=blue,
  urlcolor=magenta,
}

% \title{Automatically Detecting Specious Configuration \\ in Large Systems with Symbolic Execution}
\title{Automated Reasoning and Detection of Specious Configuration \\ in Large Systems with Symbolic Execution}
% \title{An Analytical Approach to Detect Specious Configuration \\ in Large Systems with Symbolic Execution}
\author{\rm{Yigong Hu} \\
  Johns Hopkins University
\and
\rm{Gongqi Huang} \\
  Johns Hopkins University
\\
 \addlinespace[0.5cm]
 {
\rm
Technical Report\thanks{A shorter version of this paper appears in the Proceedings of the
14th USENIX Symposium on Operating Systems Design and Implementation (OSDI '20), November, 2020}
 }
\and
\rm{Peng Huang} \\
  Johns Hopkins University 
}

\maketitle 

\input{section/abstract}
\input{section/intro}
\input{section/study}
\input{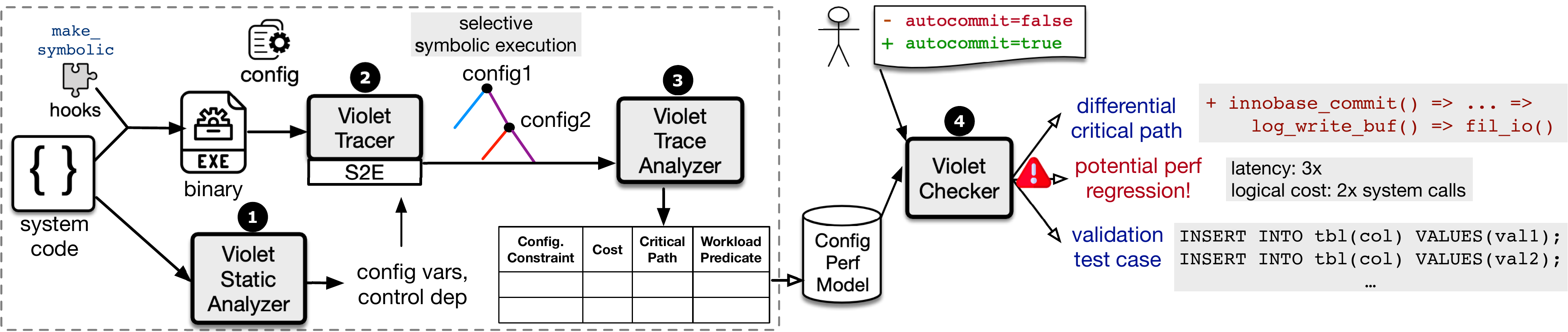}
\input{section/design}

\input{section/implement}
\input{section/evaluation}
\input{section/discuss}
\input{section/related}
\input{section/conclusion}
\input{section/ack}

{\footnotesize 
\bibliographystyle{abbrv-customized}
\bibliography{bib/reference,bib/config,bib/cloud}
}

% \clearpage 
% \begin{appendices}
%   \input{section/appendix}
% \end{appendices}

\end{document}

%% file: packages.tex
\usepackage{epsfig,endnotes}
\usepackage[T1]{fontenc}
\usepackage[scaled=0.75]{beramono}
\usepackage{amsmath,amsthm,amssymb}
\usepackage{mathptmx}
\usepackage[ruled, noend, vlined]{algorithm2e}
\usepackage{listings}
\usepackage{xcolor}
\usepackage{colortbl}
\usepackage{balance}
\usepackage{graphicx}
\usepackage{subcaption}
\usepackage{wrapfig}
\usepackage[font={small}]{caption}
\usepackage{booktabs}
\usepackage[export]{adjustbox}
\usepackage{datetime}
\usepackage{enumitem}
\usepackage{multirow}
\usepackage[binary-units=true]{siunitx}
\usepackage{pifont}
\usepackage{tikz}
\usetikzlibrary{tikzmark,calc}
\usepackage{cleveref}

%% file: preamble.tex
\DeclareMathAlphabet{\mathcal}{OMS}{cmsy}{m}{n}

\newcommand{\badconfig}{specious configuration\xspace}
\newcommand{\boldhdr}[1]{\vspace{0.05in}\noindent{\textbf{#1.}}}
\newcommand{\sboldhdr}[2]{\vspace{#1}\noindent{\textbf{#2.}}}
\newcommand{\parhdr}[1]{\vspace{0.05in}\noindent{\textbf{#1}}\hspace{0.05in}}
\newcommand{\sys}{\mbox{Violet}\xspace}

\newcommand{\sse}{S$^2$E\xspace}

\newcommand{\cmark}{\ding{51}}
\newcommand{\xmark}{\ding{55}}

\definecolor{ngray}{RGB}{102,102,102}

\newcolumntype{R}[2]{%
  >{\adjustbox{angle=#1,lap=\width-(#2)}\bgroup}%
  l%
  <{\egroup}%
}

\definecolor{bg}{rgb}{0.95,0.95,0.95}

\crefname{section}{\S}{\S\S}
\crefname{subsection}{\S}{\S\S}
\crefformat{section}{\S#2#1#3}
\crefformat{subsection}{\S#2#1#3}

\newcounter{magicrownumbers}
\newcommand\rownumber{\stepcounter{magicrownumbers}\arabic{magicrownumbers}}

%% file: section/abstract.tex
\begin{abstract}
Misconfiguration is a major cause of system failures. Prior solutions 
focus on detecting \emph{invalid} settings that are introduced by 
user mistakes. But another type of misconfiguration that continues to haunt 
production services is \emph{\badconfig}---settings that 
are valid but lead to unexpectedly poor performance in production. 
Such misconfigurations are subtle, so even careful administrators may fail to foresee them.

We propose a tool called \sys to detect such misconfiguration. We realize the crux of 
\badconfig is that it causes some slow code path to be executed, but 
the bad performance effect cannot always be triggered. \sys thus takes a novel approach that uses selective
symbolic execution to systematically \emph{reason about} the performance effect 
of configuration parameters, their combination effect, and the relationship with input. 
\sys outputs a performance impact model for the automatic detection of poor configuration 
settings. We applied \sys on four large systems. To evaluate the effectiveness 
of \sys, we collect 17 \emph{real-world} \badconfig cases. \sys detects 15 of 
them. \sys also identifies 9 unknown {\badconfig}s.
\end{abstract}

%% file: section/intro.tex
\vspace{-0.05in}
\section{Introduction}
\label{sec:intro}
\vspace{-0.05in}
Software is increasingly customizable. A mature program 
typically exposes hundreds of parameters for
users to control scheduling, caching, \emph{etc}.
With such high customizability, it is difficult to properly configure a system today,
even for trained administrators. Indeed, numerous studies and real-world failures 
have repeatedly shown that misconfiguration is a major cause 
of production system failures~\cite{Gray1986,Oppenheimer2003,YinSOSP2011,RabkinIEEE2013}. 

The severity of the misconfiguration problem has motivated solutions to detect~\cite{YuanATC2011,ZhangASPLOS2014,HuangEuroSys2015}, 
test~\cite{KellerDSN2008,Xu2013SOSP}, diagnose~\cite{Wang2003LISA,Wang2004OSDI,Whitaker2004OSDI,Attariyan2010,AttariyanOSDI2012} 
and fix~\cite{Su2007SOSP,RongATC203261,Kushman2010OSDI} misconfiguration. While these efforts help reduce misconfiguration, 
the problem remains vexing~\cite{facebook_sept_2010,aws_april_2011,aws_october_2012,
aws_december_2012,azure_december_2012,azure_february_2013,google_april_2013,twilio_july_2013,
google_jan_2014,google_cloud_april_2016,aws_feb_2017,cisco_august_2017}.
They focus on catching \emph{invalid} settings introduced due to user mistakes. % But misconfiguration is more than invalid settings. 
But another type of misconfiguration that haunts production systems, yet
not well addressed, is valid but poor configuration. For simplicity, we call them 
{\bf specious configuration}.

\begin{figure}[t]
\centering
\includegraphics[width=2.5in]{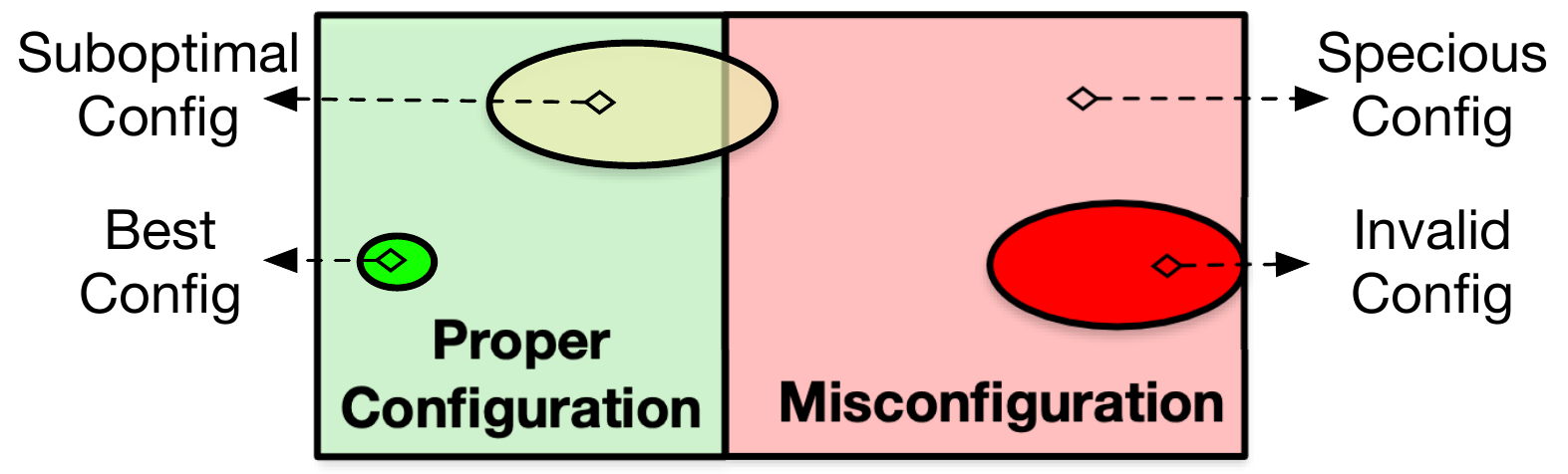}
\caption{Value space of a configuration}
\label{fig:category}
\end{figure}

Specious configuration has a broad scope. In this paper, we focus on---and 
use the term to refer to---valid settings that lead to extremely poor {\it performance}, 
which is a common manifestation in production incidents. This scope of focus is 
different from suboptimal configuration (Figure~\ref{fig:category}). The latter happens 
when a setting does not yield the best performance, but the performance is
still acceptable. 
This scope is also complementary to efforts on automated configuration 
performance tuning~\cite{SmartConf2018,BestConfig2017,CDBTune2019,Herodotou11starfish}
to search for the best setting.

Take a real-world \badconfig that caused a service outage as an example. An
engineer changed the request tracing code from a hard-coded policy
(always tracing) to be configurable with a tracing \texttt{rate} parameter.
This \texttt{rate} parameter was initially set to 0.0. To retain the same
tracing behavior as before, he decided to change the parameter to 1.0. Based on
his understanding, this change will turn on the tracing for all \emph{message}
requests that come from \emph{internal} users. But unfortunately, there was a subtle
caveat in the code that caused the actual effect to be turning on tracing for
{\it all} requests from {\it all users}, which quickly overloaded all web
servers as well as the backend databases, leading to a catastrophic 
service outage. Interestingly, before rolling out this specious configuration
to production, the change in fact went through a canary phase on a small-scale
testing cluster, which unfortunately did not manifest dramatic failure
symptoms.

Empirical evidence suggests that \badconfig like the above is prevalent. Yin et
al.~\cite{YinSOSP2011} shows that misconfiguration in the form of legal
parameters has similar or higher percentage than illegal parameters.  Facebook
reports~\cite{TangSOSP2015} that more than half of the misconfiguration in
their high-impact incidents during a three-month period are subtle, ``valid''
configurations. A recent study~\cite{SmartConf2018} on performance
configurations in distributed systems makes a similar finding. 

To reduce specious-configuration-induced incidents, we need to proactively detect it before 
production. But what makes detecting specious configuration subtle is that the parameter 
value is not a unconditionally poor choice. Rather, it is only problematic under certain
settings of some other parameters, input, and/or environment. 
Currently, administrators either informally estimate the impact based on their experience, 
or experimentally measure it by black-box testing of the program with configuration. 
However, neither of them are sufficient to reliably capture the pitfalls. 

Through analyzing real-world cases (Section~\ref{sec:study}), we realize that the 
crux of \badconfig lies in the fact that 
some slow code paths in the program or library get executed; 
but this effect can be only triggered with certain input, other configurations, and
environment. Therefore, we argue that analytical approaches are needed to
\emph{reason about} the configuration settings' performance implications under 
a variety of conditions. We propose a novel analytical tool called \textsc{\sys} 
that uses symbolic execution~\cite{KingCACM1976,CadarKLEE2008} to analyze the 
performance effect of configuration at the code level.

The basic idea of \sys is to systematically explore the system code paths with symbolic configuration
and input, identify the constraints that decide whether a path gets executed or not, 
and analytically compare different execution paths that are explored. \sys derives a configuration performance
impact model as its analysis output. A \sys checker leverages this model to
contiguously catch specious configuration in the field.
Making this basic idea work for large system software faces several challenges, 
including the intricate dependency among different parameters, the efficiency
of symbolic execution for performance analysis, complex input structure, and 
path explosion problems. \sys leverages program analysis and selective symbolic
execution~\cite{ChipounovASPLOS2011} to address these challenges.

We implement a prototype of the \sys toolchain, with its core tracer as
plugins on \sse platform~\cite{ChipounovASPLOS2011}, the static analyzer on
LLVM~\cite{Lattner2004CGO}, and the trace analyzer and checker as standalone tools. 
We successfully apply \sys on four large 
systems, MySQL, PostgreSQL, Apache and Squid. \sys
derives performance impact models for 471 parameters. To evaluate the effectiveness 
of \sys, we collect 17 real-world \badconfig cases. \sys detects 15 cases. 
In addition, \sys exposes 9 unknown \badconfig, 7 of which are confirmed by 
developers. 

In summary, this paper makes the following contributions:
\begin{itemize}[noitemsep, topsep=0pt, partopsep=0pt, leftmargin=*]
\item An analytical approach to detect \badconfig using symbolic execution and program analysis.
\item Design and implementation of an end-to-end toolchain \sys, and scaling it 
  to work on large system software.
\item Evaluation of \sys on real-world \badconfig.
\end{itemize}

The source code of \sys is publicly available available at:\\[2pt]
\phantom{x}\hspace{3em}\url{https://github.com/OrderLab/violet}

%% file: section/study.tex
\vspace{-0.05in}
\section{Background and Motivation}
\vspace{-0.05in}
\label{sec:study}
In this Section, we show a few cases of real-world \badconfig from MySQL to
motivate the problem and make the discussion concrete. 
We analyze how \badconfig affects system performance at the \emph{source code}
level. We choose MySQL because it is representative as a large system with
numerous (more than 300) parameters, many of which can be
misconfigured by users and lead to bad performance.

\subsection{Definition}
A program expects the settings of its configuration parameters to obey 
certain rules, e.g., the path exists, the min heap size does not exceed the max size. 
Invalid configurations violate those rules and usually trigger assertions or errors.

We define \badconfig to be valid settings that cause the software to
experience bad \emph{performance} when deployed to production. Admittedly, bad
performance is a qualitative criterion. We, like prior work, focus on those
less controversial issues that cause severe performance degradation.
Ultimately, only users can judge whether the performance
slowdown is sub-optimal but tolerable or it is intolerable.

Specious configuration has two classes. One is purely about performance,
\emph{e.g.}, buffer size, number of threads.  Another class is settings that
change the software functionality but the changes also have performance impact.
Both classes are important and occur in real-world systems. For the latter
class, users might want the enabled functionality and are
willing to pay for the performance cost. Thus, whether the setting is specious
or not depends on users' preferences.  Our solution addresses both forms. Its
focus is to analyze and explain the quantitative performance impact of
different settings, so that users can make better functionality-performance
trade-offs.

\vspace{-0.1in}
\subsection{Case Studies}
\label{sec:case_study}
\vspace{-0.1in}
\parhdr{\texttt{autocommit}} parameter controls the transaction commit behavior in MySQL. 
If \texttt{autocommit} is enabled, each SQL statement forms a single transaction, 
so MySQL will automatically perform a commit. 
If \texttt{autocommit} is disabled, transactions need to be explicitly 
committed with \texttt{COMMIT} statements. While \texttt{autocommit} 
offers convenience (no explicit commit required) and durability benefits, 
it also has a performance penalty since every single query will be run 
in a transaction. For some users, this performance implication may not be 
immediately apparent (especially since it is enabled by default). Even if 
users are aware of the performance trade-off, they might not know the degree 
of performance loss, only to realize the degradation is too much after deploying 
it to production. Indeed, there have been user-reported issues due to this 
setting~\cite{YinSOSP2011,slow_insert_update,rds_mysql}, and the recommended 
fix is to disable \texttt{autocommit}, and manually batch and commit 
multiple queries in one transaction.

\begin{figure}[t]
  \centering
  \captionsetup[subfigure]{aboveskip=-0.5pt,belowskip=-0.5pt}
  \begin{subfigure}[t]{0.23\textwidth}
    \includegraphics[width=\textwidth]{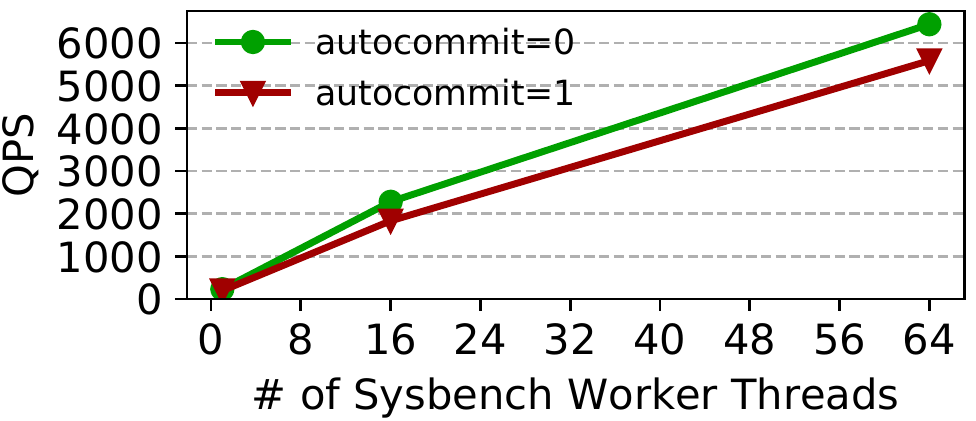}
    \caption{\footnotesize Normal workload.}
    \label{fig:autocommit_normal}
  \end{subfigure}
  \hfill
  \begin{subfigure}[t]{0.23\textwidth}
    \includegraphics[width=\textwidth]{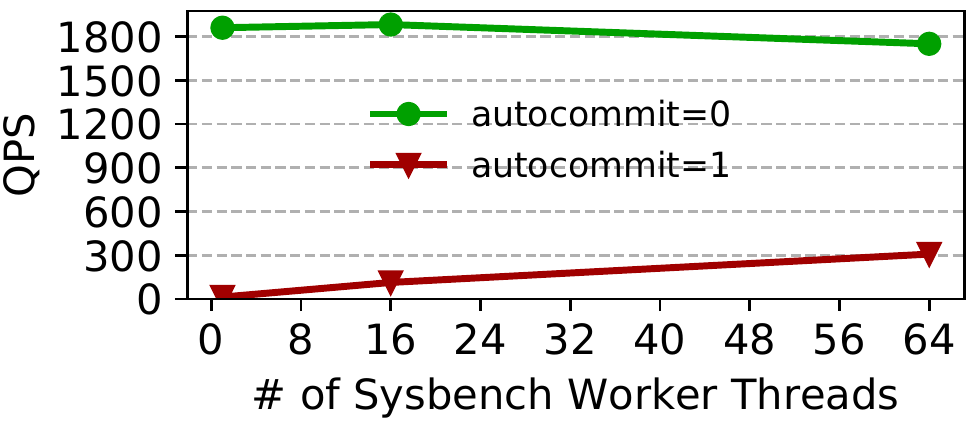}
    \caption{\footnotesize Insertion-intensive workload.}
    \label{fig:autocommit_insert}
  \end{subfigure}
  \caption{MySQL throughput for \texttt{autocommit} under two workloads. QPS: Queries Per Second.}
  \label{fig:autocommit_perf}
\end{figure}

To quantify the performance impact, we use sysbench~\cite{sysbench} to measure MySQL throughput with \texttt{autocommit} 
configuration set to be \texttt{ON} and \texttt{OFF}. The size of the database is 10 tables 
and 10K records per table. We run both a normal workload that consists of 
70\% read, 20\% write and 10\% other operations, and an insert-intensive workload. 
Figure~\ref{fig:autocommit_perf} shows the result.
We can see that in the normal workload (Figure~\ref{fig:autocommit_normal}), the performance 
difference between \texttt{ON} and \texttt{OFF} are small. But in insertion-intensive
workload (Figure~\ref{fig:autocommit_insert}), enabling \texttt{autocommit}
causes dramatically worse (6$\times$) performance.

\begin{figure}[t]
\centering
  \includegraphics[width=3.25in]{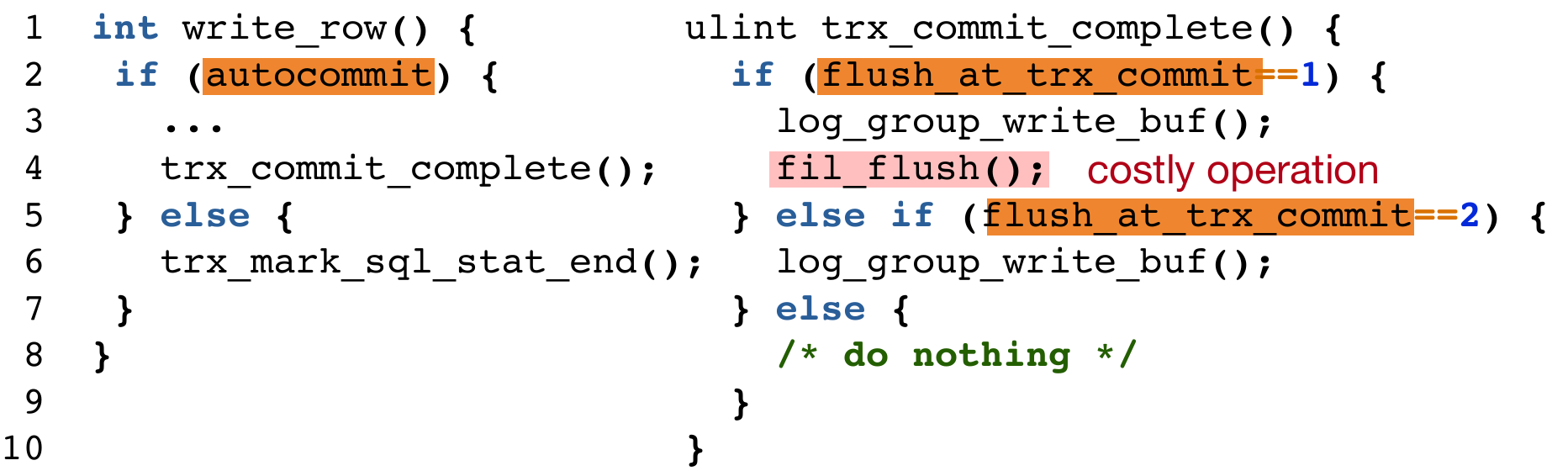}
  \caption{Simplified code snippet from MySQL related to \texttt{autocommit}. 
  The elements with orange-colored background represent configuration variables, 
  and the pink ones represent slow operations.}
\label{fig:autocommit_code}
\end{figure}

Figure~\ref{fig:autocommit_code} shows the code relevant to \texttt{autocommit}. 
We can see that \texttt{autocommit}
setting determines whether function \texttt{trx\_commit\_complete()} 
will be invoked. In this function, another parameter \texttt{flush\_at\_trx\_commit}\footnote{Its 
full name in MySQL is \texttt{innodb\_flush\_log\_at\_trx\_commit}. We abbreviate it 
and some other parameter names in this paper for readability.}
further determines which path gets executed. When that parameter is set to 1, 
compared to 2, an additional \texttt{fil\_flush} 
operation will be incurred, which has a complex logic but essentially will 
flush the table writes cached by the OS to disk through \texttt{fsync}
system call. The cost of \texttt{fsync} is the major contributor to the bad performance of 
\texttt{autocommit} mode; if \texttt{flush\_at\_trx\_commit} is 2 or 0, the performance impact 
of \texttt{autocommit} mode will be much smaller. In addition, the function in 
which \texttt{autocommit} is used---{\tt write\_row()}---is called when handling 
write type queries but not select type queries. Therefore, the performance hit 
only affects insertion/update-intensive workloads.

\begin{figure}[t]
\centering
\includegraphics[width=2.2in]{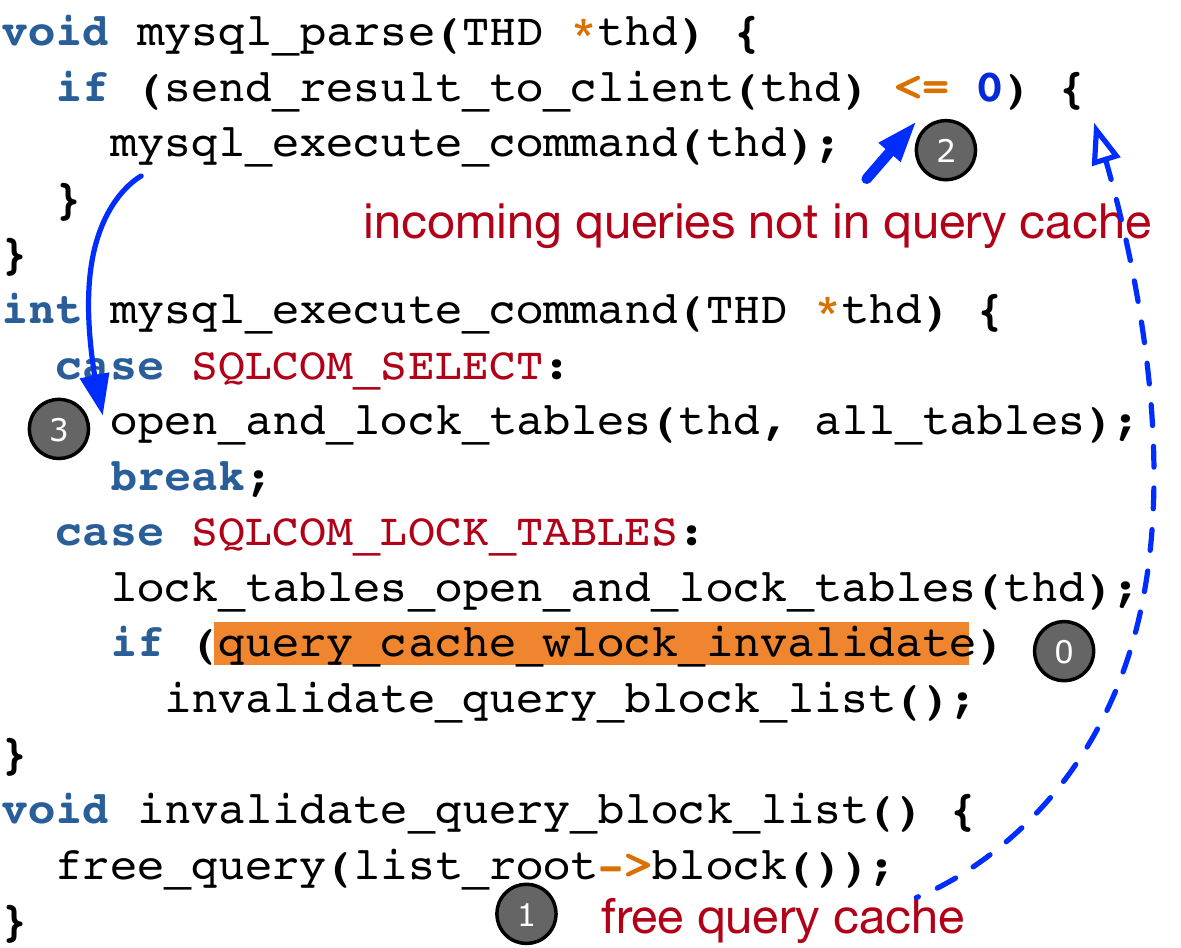}
\caption{Code affected by \texttt{query\_cache\_wlock\_invalidate}.}
\label{fig:query_cache_wlock_invalidate}
\end{figure}

\parhdr{\texttt{query\_cache\_wlock\_invalidate}} controls the validation of 
the query cache in MySQL. Normally, when one client acquires a \texttt{WRITE}
lock on a MyISAM table, other clients are \emph{not} blocked from issuing statements 
that read from the table if the query results are present in the query cache. 
The effect of setting this parameter to 1 is that upon acquisition of a \texttt{WRITE}
lock for a table, MySQL invalidates the query cache that refers to the locked table, 
which has a performance implication. 

As Figure~\ref{fig:query_cache_wlock_invalidate} shows, enabling this parameter 
leads to the \texttt{free\_query} operation (\textcolor{ngray}{\ding{202}}).
Different from the \texttt{autocommit} case, 
this operation itself is not costly. But for other clients that attempt to access the table, 
they cannot use the associated query cache (\textcolor{ngray}{\ding{203}}),
forcing them to open the table and wait (\textcolor{ngray}{\ding{204}}) while the 
write lock is held. Therefore, the effect is additional synchronization
that decreases the system concurrency, which in turn can severely hurt the 
overall system query throughput. 

Similar to \texttt{autocommit}, the performance effect depends on the execution environment and workloads. Specifically, 
the bad performance is only manifestable on the combination of MyISAM tables, \texttt{LOCK TABLES} 
statements and other clients doing \texttt{select} type queries on the locked table.

\begin{figure}[t]
\centering
  \includegraphics[width=3.25in]{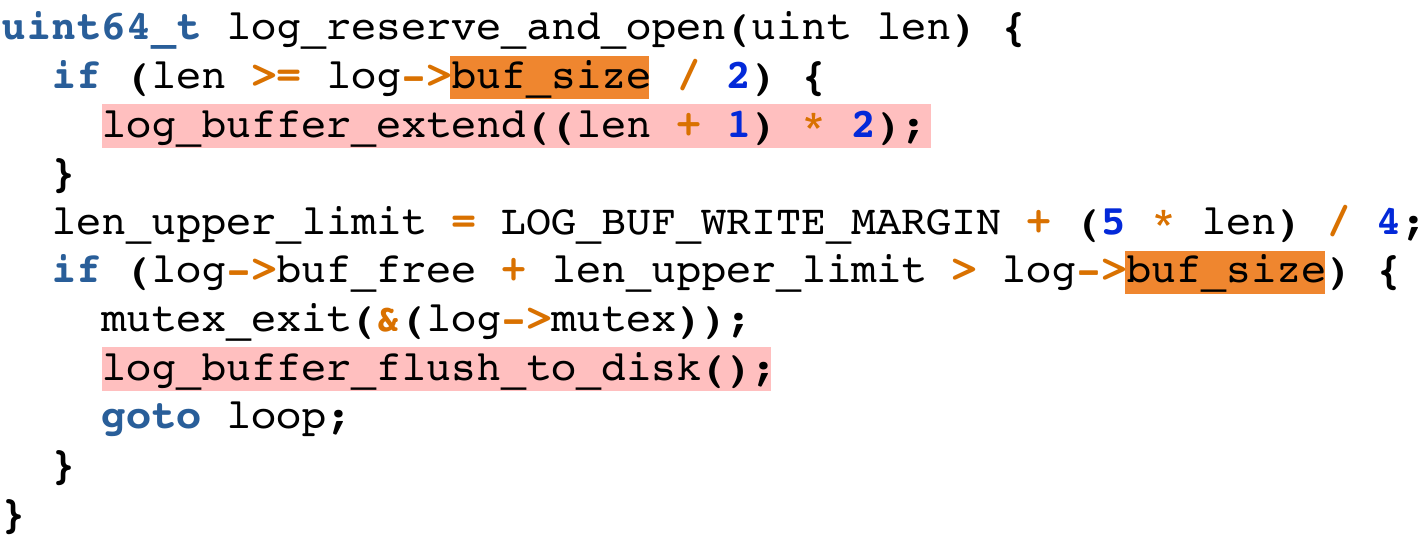}
\caption{Code affected by \texttt{innodb\_log\_buffer\_size}.}
\label{fig:innodb_log_file_size}
\end{figure}

\parhdr{\texttt{innodb\_log\_buffer\_size}} determines the size of 
the buffer for uncommitted transactions. The default 
value (8M) is usually fine. But as soon as MySQL has transactions with large blob/text
fields, the buffer can fill up very quickly and incur performance hit. As shown 
in Figure~\ref{fig:innodb_log_file_size}, the parameter setting has two possible
performance impacts: (1) if the length of a new log is larger than half of the \texttt{buf\_size}, 
the system will extend the buffer first by calling \texttt{log\_buffer\_extend}, 
which in normal cases mainly involves memory allocation. But if other 
threads are also extending the buffer, additional synchronization overhead is incurred. 
If the buffer has pending writes, they will be flushed to disk first; (2) if the 
\texttt{buf\_size} is smaller than the free size plus the length of new log, 
MySQL will trigger a costly synchronous buffer flush operation. 

\vspace{-0.08in}
\subsection{Code Patterns}
\vspace{-0.05in}
\label{sec:code_pattern}
Based on the above and other cases we analyze, we summarize four common patterns 
on how a \badconfig affects the performance of a system at the source code level:
\begin{enumerate}[noitemsep, topsep=0pt, partopsep=0pt]
  \item The parameter causes some expensive operation like the \texttt{fsync} system call to be executed.
  \item The parameter incurs additional synchronization that itself
        is not expensive but decreases system concurrency.
  \item The parameter directs the execution flow towards a slow path, \emph{e.g.}, not using cached result.
  \item The parameter triggers frequent crossings of some threshold that 
        leads to costly operations.
\end{enumerate}

The general characteristic among them is that \badconfig controls a system's execution
flows---different values cause the program or its libraries to execute
different code paths. However, the performance impact is
\emph{context-dependent}, because of the complex interplay with other
parameters, input, and environment. A \badconfig is bad when its value and
other relevant factors together direct the system to execute a path that is
significantly slower than others.

\vspace{-0.05in}
\subsection{Approaches to Detect Specious Config}
\vspace{-0.05in}
\label{subsec:approach}
Detecting \badconfig is difficult.
Operators often rely on expert experience or user manuals, which 
are neither reliable nor comprehensive. 

A more rigorous practice is to test the system together with configuration and 
quantitatively measure the end-to-end performance like throughput. 
However, if the testing does not have the appropriate input or related parameters, the 
performance issue will not be discovered. Also, because the testing is carried out in a black-box fashion,
the approach is \emph{experimental}. 
The results are tightly tied to the testing environment, 
which may not represent the production environment.
For example, in the real-world incident described in
Section~\ref{sec:intro}, that specious configuration was tested,
and the result showed 
a slight increase of logging traffic to a dependent database. But this 
increase was deemed small, so it passed the testing.

We argue that while the experimental approach is indispensable, it alone is insufficient to 
catch \badconfig. We advocate developing \emph{analytical} approaches for \emph{reasoning} 
configurations' performance effect from the system code. 
The outcome from an analytical approach includes not only a conclusion, but also answers to 
questions 
``how the parameter affects what operations get executed?'', ``what kind of input 
will perform poorly/fine?'', and ``does the effect depend on other parameters?''. 
In addition, the analysis should enable \emph{extrapolation} to different contexts, 
so users can project the outcome with respect to specific workload or environment. 

A potential analytical approach is static analysis. Indeed, the code patterns 
summarized in Section~\ref{sec:code_pattern} can be leveraged to detect potential 
\badconfig. However, 
mapping them at concrete code construct level requires substantial 
domain knowledge. Also, the performance effect involves many complex 
factors that are difficult to be deduced by pure static analysis.

The observations in Section~\ref{sec:code_pattern} lead us to realize that
the crux is some slow path being conditionally executed. Thus, we can transform 
the problem of detecting \badconfig to the problem of finding slow execution 
flow plus deducing the triggering conditions of the slow execution. 

%% file: section/overview.tex
\section{Overview of \sys}
\label{sec:overview}
We propose an analytical approach for detecting \badconfig, and 
design a tool called \textsc{\sys}. 
\sys aims to comprehensively reason about the performance effect of system 
configurations: (1) explore the system without being 
limited by particular input; (2) analyze the performance effect without being 
too tied to the execution environment. 

\begin{figure*}[t]
  \centering
  \includegraphics[width=5.25in]{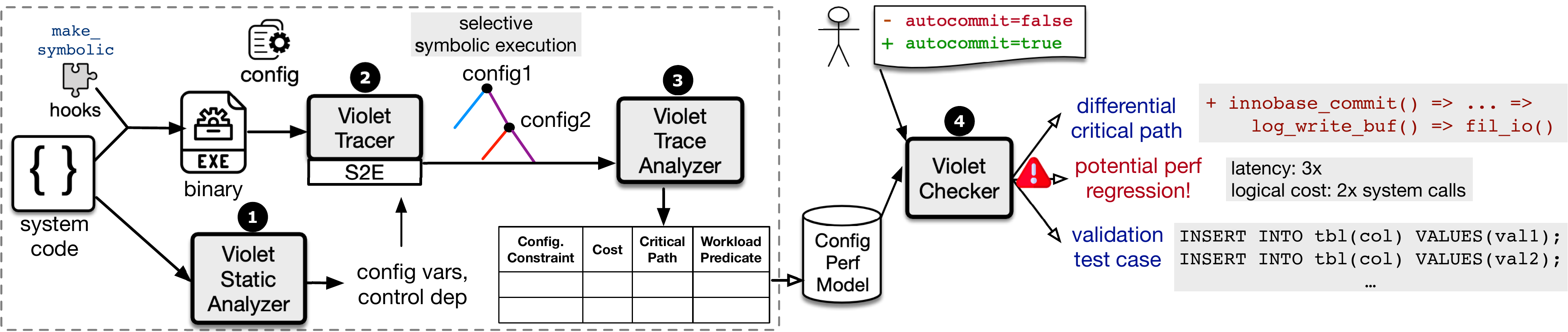}
  \caption{Overview of \sys.}
  \label{fig:overview}
  \vspace{-0.15in}
\end{figure*}

Our insight is that the subtle performance effect of a specious parameter is 
usually reflected in different \emph{code paths} getting executed, depending 
on \emph{conditions} involving the parameter, input and other parameters, 
and these paths have significant \emph{relative} performance differences. 
Based on this insight, \sys uses symbolic execution with assistance of static 
analysis to thoroughly explore the influence of configuration parameters on 
program execution paths, identify the conditions leading to each execution, 
and compare the performance costs along different paths. After these analyses, 
\sys derives a configuration performance impact model that describes the 
relationship between the performance effect and related conditions. 
In this Section, we give an overview of \sys 
(Figure~\ref{fig:overview}). We describe the design of \sys in Section~\ref{sec:design}.

\subsection{Symbolic Execution to Analyze Performance Effect of Configurations}
\label{sec:se_background}
\boldhdr{Background} Symbolic execution~\cite{KingCACM1976,CadarKLEE2008} is 
a popular technique that systematically explores a program. 
Different from testing that exercises a single path of the 
program with concrete input, symbolic execution explores multiple paths 
of the program with symbolic input and memorizes the \emph{path constraints} 
during its exploration. When a path of interest (e.g., with
\texttt{abort()}) is encountered, the execution engine generates 
an input that satisfies the constraint, which can be used as a test case. 
Compared to random testing, symbolic execution systematically explores 
possible program paths while avoiding redundancy. Consider this snippet: 
\vspace{-0.1in}
\begin{center}
  \includegraphics[width=1in]{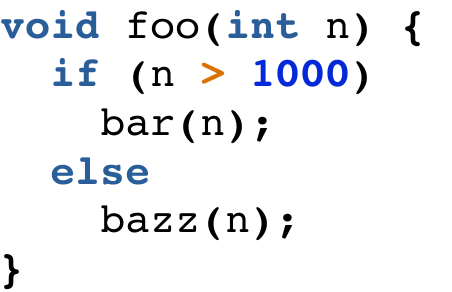}
\end{center}
\vspace{-0.1in}
Testing may blindly test the program many times with different values of \texttt{n}, e.g., 
1, 10, 20, etc., but they all exercise the same path without triggering
the call to \texttt{bar()}. In comparison, if we use symbolic execution: \\
\includegraphics[width=3in]{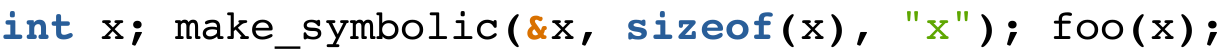}\\
we can explore the two paths of \texttt{foo} 
by deriving only two concrete values of \texttt{n} to satisfy the path constraints.

\boldhdr{Basic Idea} Configuration is essentially one type of external 
input to a program. The basic idea of \sys is simple---make the 
parameters of a system symbolic, measure the cost along each execution path explored, 
and comparatively analyze the costs. The path constraints that the symbolic execution 
engine memorizes characterize the conditions about whether and when a parameter 
setting is potentially poor. Take Figure~\ref{fig:autocommit_code}
as an example. \sys makes variable \texttt{autocommit} symbolic. Function
\texttt{write\_row} will fork at line 2. The first path goes into the \texttt{if} branch, with a constraint \texttt{autocommit == 1}. 
When \texttt{trx\_commit\_complete} is called in 
the first path, it encounters another parameter \texttt{flush\_at\_trx\_commit},
which is also made symbolic. Two additional paths are forked within 
that function. While exploring these paths, \sys records 
a set of performance cost metrics.

Since the subtle performance effect of \badconfig is often only triggered
under specific input, besides configuration parameters, \sys can also make the 
input symbolic. For the example in Figure~\ref{fig:autocommit_code},
the input will determine whether the \texttt{write\_row} function will 
be called or not.
Only \texttt{insert} type queries will invoke \texttt{write\_row}. 
This input constraint will be recorded so the analysis later can identify
what \emph{class} of input can trigger the \badconfig.

\subsection{Violet Workflow}
Figure~\ref{fig:overview} shows the workflow of \sys. 
The input to \sys is system code. We require source code to identify 
the program variables corresponding to configuration parameters. In addition, 
as we discuss later (Section~\ref{sec:control_dep}), \sys uses static analysis 
to assist the discovery of dependent parameters. 
To symbolically execute the target system, we leverage a state-of-the-art
symbolic execution platform S$^2$E~\cite{ChipounovASPLOS2011} and insert 
hooks to \sse APIs in the system code to make parameters and input symbolic.
We design the \sys execution tracer as \sse plugins to analyze the performance 
during state exploration and write the results to a trace. 
The \sys trace analyzer conducts comparative cost analysis, differential critical 
path analysis, etc. The outcome is a configuration performance impact model that 
describes the relationship among configuration constraints, cost, critical path, 
and input predicate. 

\sys further provides a checker to deploy with the software at user sites. 
The checker consumes the constructed configuration impact model to continuously detect whether a 
user-site configuration file or update can potentially lead to poor 
performance. Upon the detection of potential \badconfig, the \sys checker 
reports not only the absolute performance result, but also the logical cost and 
critical path to explain the danger. The checker also outputs a validation
test case based on the input predict that provides hints to users about what input 
can expose the potential performance issue.

%% file: section/design.tex
\section{The Design of \sys}
\label{sec:design}
\vspace{-0.05in}

In this Section, we describe the \sys design (Figure~\ref{fig:overview}). 
We need to address several design challenges. First, configurations have intricate 
dependencies among themselves and with the input, but making 
all of them symbolic easily leads to state space explosion. Second, 
conducting performance analysis in symbolic execution is demanding due to 
lack of explicit assertion point, mixed costs, overhead, etc. Third, 
deriving performance model from code requires balance between being generalizable 
(not too tailored to specific input or environment) and being realistic (reflects 
costs in real executions).

\vspace{-0.05in}
\subsection{Make Config Variable Symbolic}
\vspace{-0.05in}
\label{sec:var_symbolic}
The starting point for \sys is to make parameters symbolic. 
A na\"ive way is to 
make the entire configuration file a symbolic blob. While this approach is transparent 
to the target program, it easily leads to path explosion even at the program 
initialization stage. 
An improvement could be only making the configuration value string symbolic during parsing. 
e.g., \\
\includegraphics[width=3in]{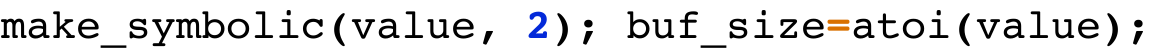}\\
But the execution would still spend significant time in the parsing (\texttt{atoi}). 
Also the parameter value range will be limited by the string size. 

\begin{figure}[t]
\centering
  \includegraphics[width=3.2in]{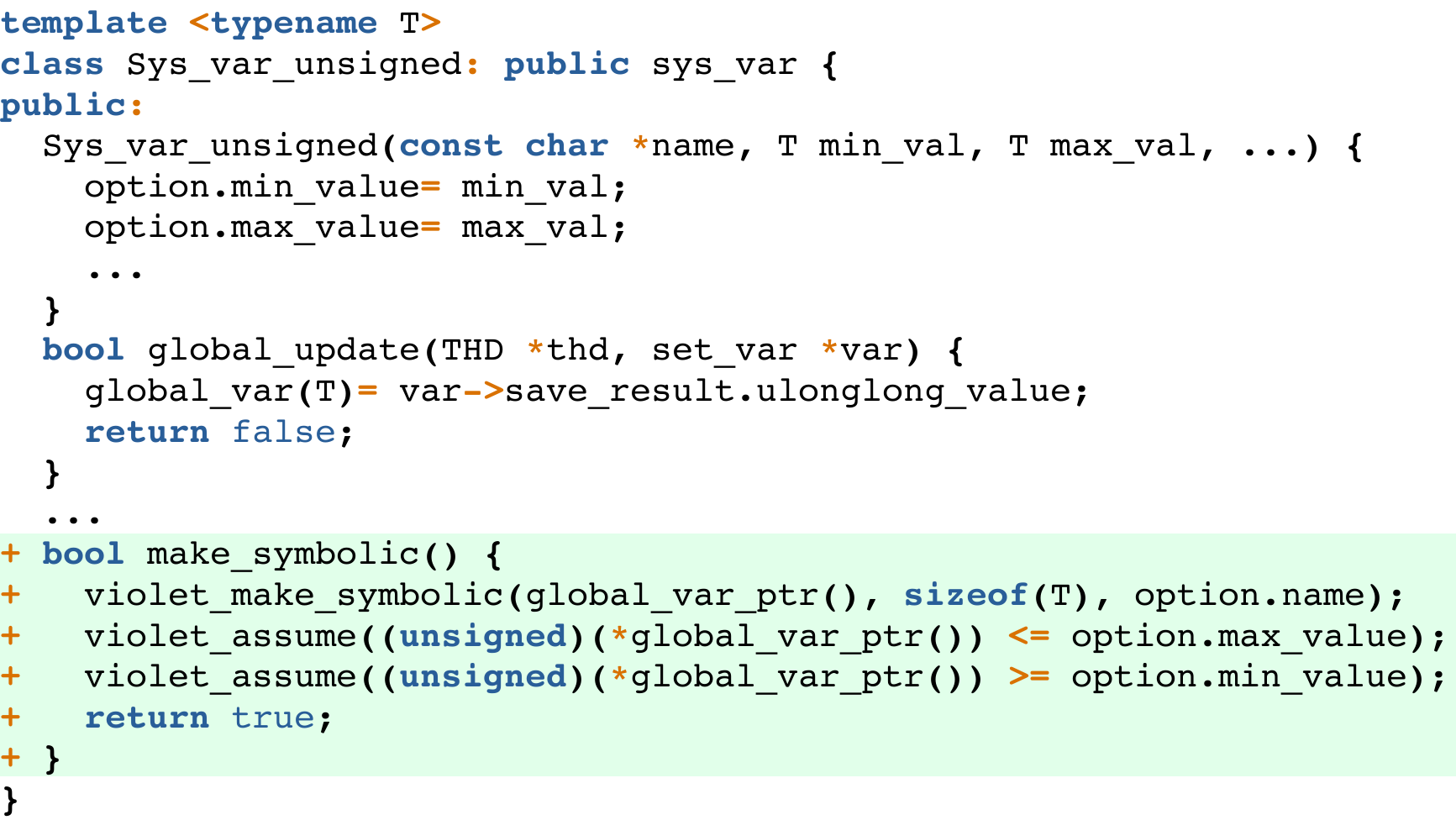}
  \caption{Add API to one config. data structure in MySQL.}
  \label{fig:config_hook}
\end{figure}

We should identify the program variables that store configuration parameters and 
directly make these variables symbolic. Prior works~\cite{Xu2013SOSP,PCheckOSDI16} 
observe that the mature software typically uses uniform interfaces such as an array of \texttt{struct} to store 
parameters. 
Thus they annotate these interfaces to extract variable mappings in static analysis. 
For \sys, we need to additionally identify the parameter type and value constraints 
defined by the program (e.g., \numrange{1}{10}) to
restrict the symbolic value. This is because we are only interested in exploring 
the performance effect of \emph{valid} values. 

Since typically all the config variables are readily accessible after some
point 
during initialization, we take a simple but accurate approach: insert a hook
function directly in the source code right after the parsing function and 
programmatically enumerates these variables and make them symbolic
using their type and other info. In this hook function, we read an external
environment variable \texttt{VIO\_SYM\_CONFIGS} to decide which target
parameter(s) to make symbolic. 

\begin{figure}[t]
\centering
  \includegraphics[width=3.2in]{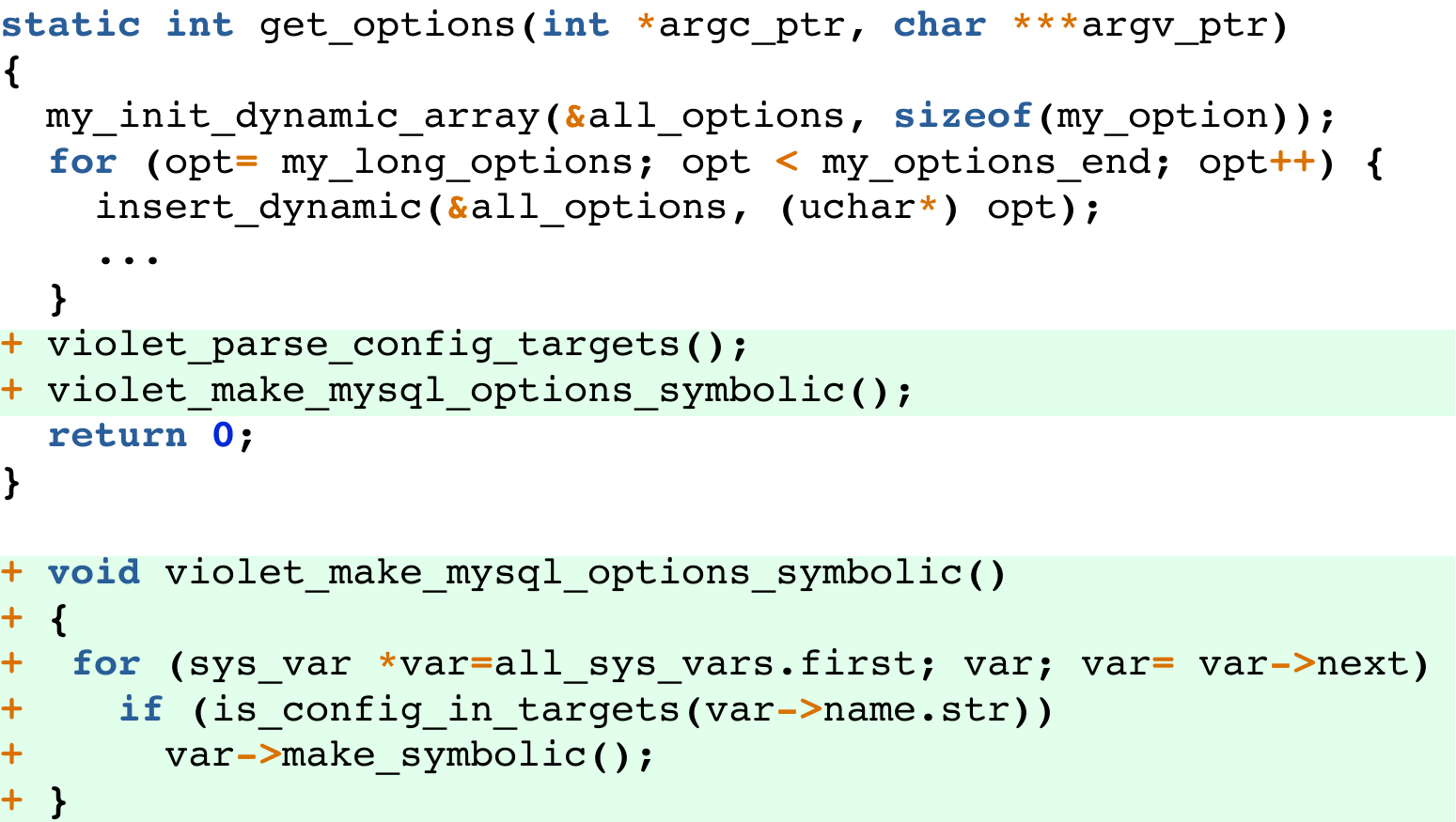}
  \caption{Call symbolic hooks after config. parsing in MySQL.}
  \label{fig:call_config_hook}
\end{figure}

Take MySQL as an example. Its configuration parameters are represented by a number 
of \texttt{Sys\_var\_*} data structures in the code, depending on the parameter's 
type. We add a \texttt{make\_symbolic} API to these data structures, which uses 
the type, name, value range information to call the Violet library to make the 
backing store symbolic. Figure~\ref{fig:config_hook} shows an example of the 
added hook API. Then after MySQL finishes parsing its configurations, we 
iterate through all configuration variables (Figure~\ref{fig:call_config_hook}), 
which are stored in a global linked list called \texttt{all\_sys\_vars}. If the 
parameter is in the target set, we invoke its new \texttt{make\_symbolic} API.

\begin{figure}[t]
\centering
  \includegraphics[width=3.0in]{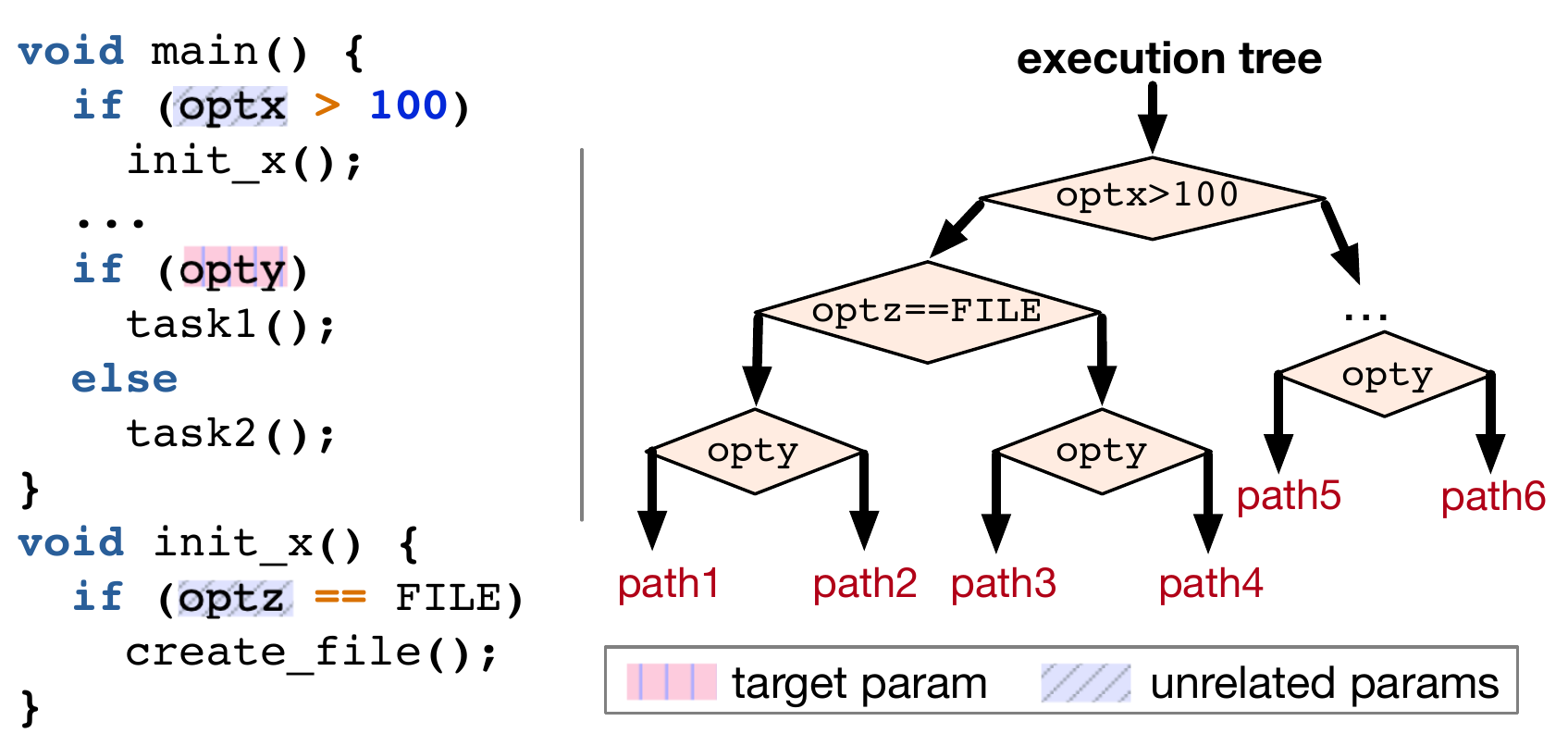}
  \caption{Making unrelated parameters symbolic results in excessive state explorations and confusing conclusions.}
  \label{fig:control_dep_example1}
  \vspace{-0.11in}
\end{figure}

\vspace{-0.05in}
\subsection{Make Related Config Symbolic}
\label{sec:config_set}
The performance effect of a parameter usually depends on the values of other
parameters. Thus, if we only make one parameter symbolic while leaving other
parameters concrete, we will only explore incomplete execution paths and
potentially miss some problematic combination that leads to bad performance. A
straightforward solution is to make all parameters symbolic. Since symbolic
execution only forks if a symbolic value is used branch conditions, this
approach seems to be feasible. However, the problem with this approach is that 
most combinations of configuration parameters are unrelated but will be 
explored during symbolic execution.

Figure~\ref{fig:control_dep_example1} illustrates the problem. 
Suppose we are interested in the performance effect of \texttt{opty}. 
If we simply make all parameters (\texttt{optx}, \texttt{opty}, 
\texttt{optz}) symbolic in hope of exploring the combination 
effect, there will be at least 6 execution paths being explored. But 
\texttt{opty} is unrelated to \texttt{optx} and \texttt{optz}.
The performance impact of \texttt{opty} is only 
determined by the cost of its branches. 
For large programs, the target parameter could be used deep in the code. 
Including unrelated parameters in the symbolic set can cause the symbolic execution 
to waste significant time
or get stuck before reaching the interesting code place to explore
the target parameter. The analysis result can also cause confusions. 
For example, it might suggest only when \texttt{optx>100 \&\& optz==FILE \&\& opty} 
is true will there be a performance issue and miss detecting \badconfig 
when \texttt{opty} is true but \texttt{optx <= 100} or \texttt{optz != FILE}.

Therefore, instead of making all parameters symbolic at once, we carefully
choose the set of parameters to symbolically execute together. In particular,
related parameters are usually control dependent on each other. We discover the
parameter control dependency with methods described in the following Section.

\vspace{-0.1in}
\subsection{Discover Control Dependent Configs}
\label{sec:control_dep}
\begin{figure}[t]
  \centering
  \includegraphics[width=3.25in]{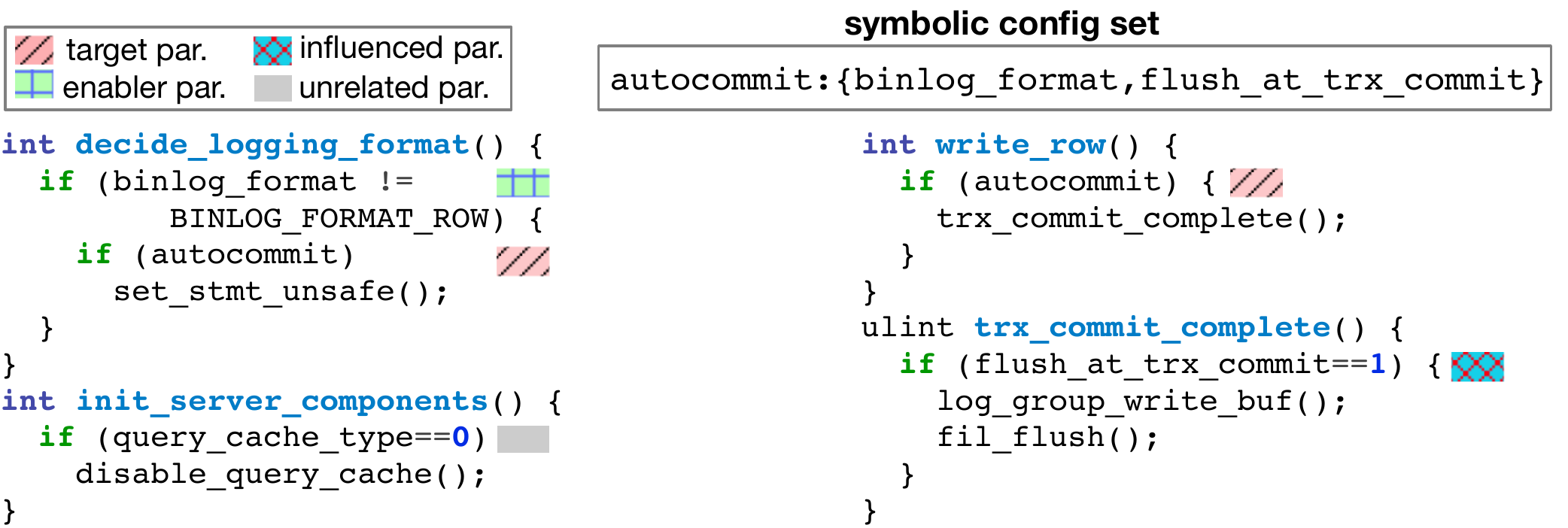}
  \caption{Symbolic config set based on control dependencies.}
  \label{fig:control_dep_set}
  \vspace{-0.05in}
\end{figure}

\sys statically analyzes the control dependency relationship of parameters to 
determine a reduced symbolic \emph{parameter set}. The static analysis result can 
significantly help mitigate the path exploration problem during symbolic execution 
phase.

For a target parameter $C$, \sys identifies two kinds
of related parameters to put in its symbolic set. The \emph{enabler parameters}
are those that $C$ is control dependent on. The \emph{influenced
parameters} are those that are control dependent on $C$. 
Figure~\ref{fig:control_dep_set} shows an example. For target parameter
\texttt{autocommit}, it is used in \texttt{decide\_logging\_format} and \texttt{write\_row},
it has an enabler parameter \texttt{binlog\_format}, which decides 
if \texttt{autocommit} will be activated. \texttt{autocommit} 
itself influences the performance effect of parameter \texttt{flush\_at\_trx\_commit}.
Thus, for \texttt{autocommit}, the set of related parameters to make symbolic together is 
$\{\texttt{binlog\_format},\texttt{flush\_at\_trx\_commit}\}$.

Informally, program element $Y$ is control dependent on element $X$ if whether $Y$'s 
executed depends on a test at $X$. More formally, control dependency is captured 
by postdominator relationship in program Control Flow Graph (CFG).
Node $b$ in the CFG postdominates node $a$ if every path from $a$ to the 
exit node contains $b$. $Y$ is control dependent on $X$ if there is a path 
$X \rightarrow Z_1 \rightarrow \ldots \rightarrow Z_n \rightarrow Y$ such that 
$Y$ postdominates all $Z_i$ and $Y$ does \emph{not} postdominate $X$. 
We use postdominator as a building block for our analysis. 
But our notion of control dependency is broader than the classic definition. For example, 
in the following code snippet (1)
\begin{center}
\includegraphics[width=2in]{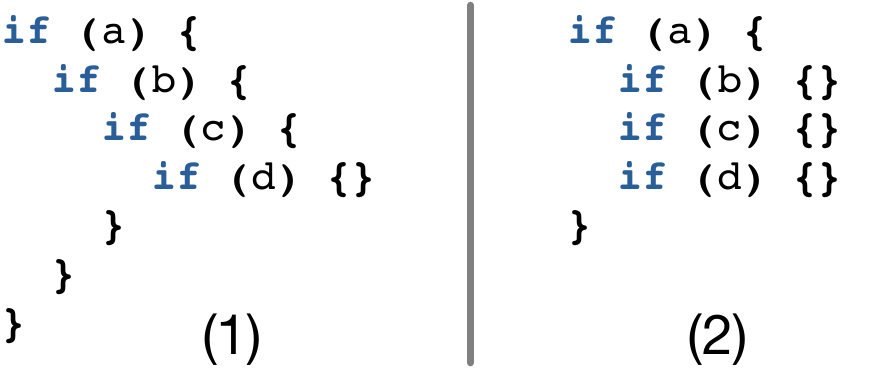}
\end{center}
, the classic definition does \emph{not} regard \texttt{a} and \texttt{d}
as being control-dependent (\texttt{c} and \texttt{d} are, and the \texttt{a} 
and \texttt{d} in (2) are).
But for us, all the four parameters are control dependent. 

\input{section/control_dep_algo}

Our analysis is divided into two steps (Algorithm~\ref{algo:related_config}). 
The first step computes the enabler parameters. % 
Algorithm~\ref{algo:enabler_config} lists the core logic for this step. 
\sys builds a call graph of the program. For target parameter $p$, 
it locates the usage points of $p$ and extracts the call 
chains starting from the entry function to the function $f$ that encloses 
a usage point. If any caller $g$ in the call chain uses some other parameter $q$, 
we check if the callsite in $g$ that eventually reaches $f$ is 
control dependent on the usage point of parameter $q$ in $g$. If so,
$q$ is added to the enabler parameter set of $p$. \sys identifies
enabler parameters within $f$ through intra-procedural control dependency.

In the second step, \sys calculates the influenced parameters from
the computed enabler parameter sets of all parameters. The
related config set is a union of the influenced set and enabler set (Algorithm~\ref{algo:related_config}). 
We also capture control dependency that involves simple data flow. 
For example, 
\includegraphics[width=3.2in]{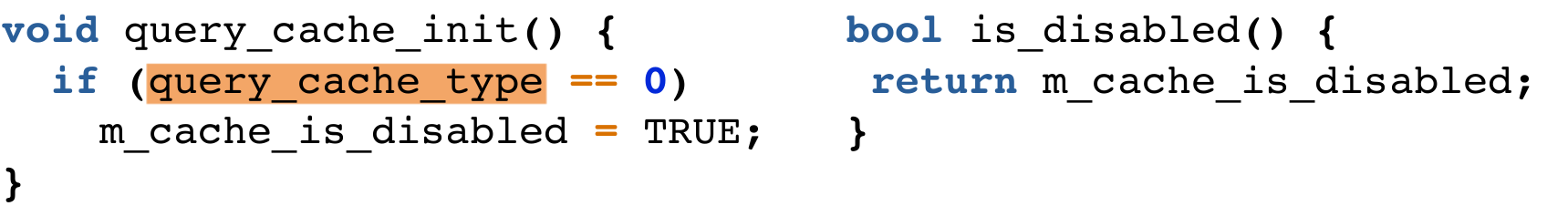}
\\ 
any parameter that is control dependent on the regular variable \texttt{m\_cache\_is\_disabled} 
or return value of \texttt{is\_disabled()} is also considered to be related 
to parameter \texttt{query\_cache\_type}.

The static analysis result can be inaccurate due to imprecision 
in the pointer analysis, call graph, infeasible path problem, etc.
Our general principle is to be conservative and over-approximate the 
set of related parameters for a target parameter. During symbolic execution, 
having a few false control dependent parameters does not greatly
affect the performance or analysis conclusion and they can manifest
through the symbolic execution log if they do cause serious issues.

\vspace{-0.1in}
\subsection{Execute Software Symbolically}
After the target software is instrumented with the symbolic execution hooks, 
\sys symbolically executes the software with a concrete configuration
file. The hook function reads the \texttt{VIO\_SYM\_CONFIGS} environment
variable and makes symbolic the program variables corresponding to the specified 
parameter. In addition, the function parses the control dependency analysis (Section~\ref{sec:control_dep}) 
result file and makes variables in the related parameter set symbolic as well. 
Other parameters' program variables get the concrete values from the
configuration file. Besides parameters, \sys can also make program input symbolic 
to explore its influence on the configuration's performance impact. This is done 
through either symbolic arguments (\texttt{sym-args}) or identifying 
the input program variables and inserting \texttt{make\_symbolic} calls in the code.

\vspace{-0.1in}
\subsection{Profile Execution Paths}
\label{sec:profile_path}
\begin{figure}[t]
\centering
  \includegraphics[width=3.2in]{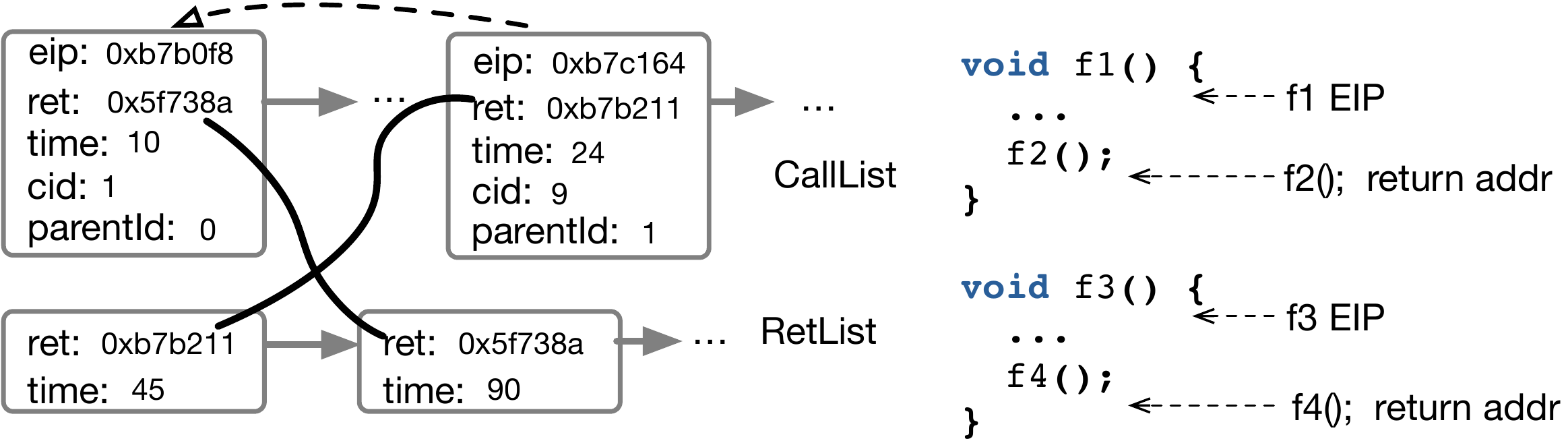}
  \caption{Match call/return records.}
  \label{fig:profile_call_records}
\end{figure}

To measure the symbolic parameters' performance effect, \sys implements a
tracer on top of the symbolic execution engine, specifically as a set of
plugins on the \sse platform. 

\boldhdr{Measure Function Call Latency} We measure function call latency by capturing 
the call and return signals emitted by \sse during symbolic execution. 
To calculate the latency, a straightforward way is to maintain a stack of call record 
and pops the top element upon receiving a return signal. This algorithm assumes 
that the call/return signals are paired and the callee's return signal comes before 
the caller's. But we observe this assumption does not always hold under \sse. 
We use a safer method based on return addresses to calculate latency. 
In particular, the \sys tracer records the \texttt{EIP} register value, return 
address, and timestamp on each call and return signal. The records are stored 
in two lists. Later, the tracer matches call record list with return record
list based on return address fields (Figure~\ref{fig:profile_call_records}). 
The latency for a matched function call is the return record's timestamp minus the 
call record's timestamp. The total latency of each state (execution path) can be 
obtained from the latency of the root function call.

For multi-threaded programs, function calls from different threads can get
mixed up. To address this issue, the \sys tracer stores the current thread id
in each profile record and partitions the call and return lists by thread id.

\boldhdr{Re-Construct Call Paths} The tracer records the function call profile 
to break down total latency and to enable differential critical path analysis (\S\ref{sec:trace_analysis}).
To get the call chain relationship, instead of costly stack frame walk,
the tracer uses a simple method with little overhead that just assigns each call 
record a unique incrementing \texttt{cid}. Later, the tracer reconstructs 
the call chain by iterating through all call records in order. If 
(1) call record \emph{A}'s \texttt{cid} is larger than call record \emph{B}'s 
\texttt{cid}, (2) the return address of \emph{A} is larger than \emph{B}'s 
\texttt{EIP} (the start address of that function), and (3) the difference of the two 
addresses is smallest among all other pairs (i.e., \emph{B}'s start 
address is closest to the return address in \emph{A}), then we assign \emph{A}'s 
\texttt{parentId} to be \emph{B}'s \texttt{cid} and update the current distance. % The

\boldhdr{Measure Logical Costs}
Besides absolute latency, we also measure a set of logical cost metrics by
a similar method of capturing low-level signals from \sse. In particular, for
each execution path, we measure the number of instructions, the number of
system calls, the number of file I/O calls, the amount of I/O traffic, the
number of synchronization operations, network calls, etc.  These
logical costs are useful to surface performance issues other than just long
latency. 
They are also crucial for 
enabling extrapolation of the result to different settings. For example, if the \sys tracer finds out one execution
path has a much higher number of \texttt{write} syscalls compared to other
paths whereas their latencies are similar. This could be an artifact of the
test server having a powerful hard disk or a large buffer cache. But the software might
perform poorly in a different environment.

\input{section/cost_table_example}

The \sys tracer maintains a separate performance profile for each execution path (state)
so we can compare the performance effect of different paths. We also need to record 
the path constraints to identify the parameter combination and the \emph{class} of input that leads to 
the execution path. 
The tracer records the final path constraint when an execution path terminates 
or it exceeds some user-specified cost threshold.

\vspace{-0.1in}
\subsection{Analyze State Traces}
\label{sec:trace_analysis}
Once the symbolic execution finishes, the \sys trace analyzer parses the performance 
traces. 
It then builds a cost table. Each row represents a state (path) that was explored in symbolic execution. 
The analyzer does a pair-wise comparison of performance in different rows. If the 
performance difference ratio exceeds a threshold (default 100\%), the analyzer marks that state suspicious. The analyzer compares 
not only the absolute latency metric but also the collected logical metrics. Even if the 
latency difference does not exceed the threshold but some logical metric
does, the analyzer still marks the state.

Not all pair comparisons are equally meaningful when the symbolic execution explored multiple symbolic 
variables. To elaborate, assume our target parameter is \texttt{autocommit}, 
which has a related parameter \texttt{flush\_log}. Since both are made symbolic, 
one state could represent constraint \texttt{autocommit==0 \&\& flush\_log==1} 
and another state could represent constraint \texttt{autocommit==1 \&\& flush\_log==2}.
In this case, comparing the costs of these two states is not very meaningful. 

The analyzer tries to compare state pairs that are most ``similar'' first.
Determining the similarity of two paths can be challenging. 
We use a simple approach: in one state's constraints formula, 
for each constraint involving a related parameter, if it also appears in the other 
state's formula, the similarity count is incremented by one. This method is imprecise 
as it merely checks the appearances, not constraint equivalence. 
For our use cases, the inaccuracies are generally acceptable. Besides, the
analyzer can compare all pairs first, surface the bad state-pairs, and then 
we can decide the meaningfulness of the suspicious pairs.

For each pair that has a significant performance difference, the analyzer 
computes the \emph{differential} critical path. It first finds the longest 
common subsequence of the call chain records in the two states. Then it creates 
a diff trace that stores the common records with performance metrics subtracted, as well 
as the records that only appear in the slower state. % With the diff trace, 
The analyzer finally locates the call record (excluding entry) with the largest 
differential cost and constructs the critical call path based on the 
\texttt{cid} and \texttt{parentId} of the call records.

When \sys makes the input symbolic, the path constraints in each state will
contain constraints about the input. The analyzer separates the input 
related constraints as input predicate. This is useful to tell what 
\emph{class} of input can expose the potential performance issue for the 
combination of parameter values that satisfies the configuration constraint 
in a state. The final output from the \sys analyzer is the configuration
performance impact model that consists of the raw cost table (Table~\ref{tab:impact_example}) 
with configuration constraints, cost metrics, and input predicate for each state, 
the state pairs that have significant performance difference, and the differential 
critical paths.

\vspace{-0.1in}
\subsection{Continuous Specious Config Checker}
\label{sec:checker}
\sys provides a standalone checker tool to detect \badconfig. It leverages the
configuration performance impact model from the analyzer and validates a
concrete user configuration file. The checker tool supports three modes: 
\begin{enumerate}[noitemsep, topsep=0pt, partopsep=0pt]
    \item Some config update introduces performance regression.
    \item Some default parameter is poor for users' specific setup.
    \item Code upgrade or workload change make old setting poor.
\end{enumerate}
For scenario 1, the checker references the cost table and locates the state(s) that 
have configuration constraints satisfying the updated parameter' old value and the 
parameter's new value. If the state pair has significant performance difference, 
the checker alerts the operators and generates a test case based on 
the input predicate for operators to confirm the performance regression. 
For scenario 2, the checker validates if the state that the default value lies in appears in some poor state-pair. 
If so, it means this default value potentially performs significantly worse than 
another value. 
For scenario 3, if the system code changes, \sys rebuilds the cost impact table. The checker then 
identifies if some state in the new table performs much worse compared 
to the old cost table. If workload changes, the checker validates if
cost table rows that previously satisfy the input predicate perform
worse compared to rows that satisfy the input predicate now. 

\input{section/scale}

%% file: section/control_dep_algo.tex
\begin{algorithm}[t]
  % \SetAlgoNoLine
  \DontPrintSemicolon
  \SetKwInput{KwFunc}{Func}
  \SetInd{0.5em}{0.25em}
  \KwFunc{\texttt{GetRelatedConfigs}}
  \KwIn{$\mathcal{P}$: target program, $\mathcal{C}$: all parameter vars in $\mathcal{P}$}
  \KwOut{$\mathcal{M}$: map from each parameter in $\mathcal{C}$ to the set of related parameters}
  % \tcc{each value in \texttt{is\_map} is the influenced parameter set}
  $\mathcal{M} \gets$ \texttt{\{\}}, \texttt{es\_map} $\gets$ \texttt{\{\}}, \texttt{ins\_map} $\gets$ \texttt{\{\}}\;
  \ForEach{\texttt{p} $\in$ $\mathcal{C}$} {
    \texttt{es} $\gets$ \texttt{GetEnablerConfig(p, $\mathcal{P}$}\texttt{)}\;
    \texttt{es\_map[p] $\gets$ es}\;
    \ForEach{\texttt{q} $\in$ \texttt{es}} {
      \texttt{ins\_map[q]} $\gets$ \texttt{ins\_map[q]} $\cup$ \texttt{\{p\}}
    }
  }
  \ForEach{\texttt{p} $\in$ $\mathcal{C}$} {
    $\mathcal{M}$\texttt{[p]} $\gets$ \texttt{es\_map[p]} $\cup$ \texttt{ins\_map[p]}
  }
  \Return{$\mathcal{M}$}
  \caption{Compute related parameters}
  \label{algo:related_config}
\end{algorithm}

\begin{algorithm}[t]
  % \SetAlgoNoLine
  \DontPrintSemicolon
  \SetKwInput{KwFunc}{Func}
  \SetInd{0.5em}{0.5em}
  \KwFunc{\texttt{GetEnablerConfig}}
  \KwIn{$\mathcal{P}$: target program, \texttt{p}: target parameter variable}
  \KwOut{\texttt{ES}: set of enabler parameters for \texttt{p}}
  \texttt{ES} $\gets \varnothing$, \texttt{p\_usages} $\gets$ \texttt{getUsages(p, }$\mathcal{P}$\texttt{)}\;
  \ForEach{\texttt{p\_usage} $\in$ \texttt{p\_usages}} {
    \texttt{chains} $\gets$ \texttt{callgraph.getPaths(p\_usage.func)}\;
    \texttt{chains.append(p\_usage.func)}\;
    \ForEach{\texttt{q\_func} $\in$ \texttt{chains}} {
      \texttt{q\_usages} $\gets$ \texttt{getConfigs(q\_func)}\;
      \ForEach{\texttt{q\_usage} $\in$ \texttt{q\_usages}} {
        \If{\texttt{q\_usage} $\neq$ \texttt{p\_usage} $\wedge$ \texttt{q\_usage.var} $\neq$ \texttt{p}} {
          \If{\texttt{q\_func} $\neq$ \texttt{p\_func}} {
            \texttt{p\_site} $\gets$ \texttt{callSite(q\_func, p\_func)}
          }\Else (\tcc*[h]{within the p's usage function}){
            \texttt{p\_site} $\gets$ \texttt{p\_usage.instruction}
          }
          \If{\texttt{ctrlDep(p\_site, q\_usage)}} {
            \texttt{ES} $\gets$ \texttt{ES} $\cup$ \{\texttt{q\_usage.param}\}
          }
        }
      }
    }
%     \texttt{q\_usages} $\gets$ \texttt{getUsage(p\_func)}\;
%     \ForEach{\texttt{q\_usage} $\in$ \texttt{q\_usages}} {
%       \If{\texttt{p} $\neq$ \texttt{q\_usage.var}} {
%         \If{\texttt{ctrlDep(p\_usage, q\_usage)}} {
%           \texttt{ES} $\gets$ \texttt{ES} $\cup$ \{\texttt{q\_usage.param}\}
%         }
%       }
%     }
  }
  \Return{\texttt{ES}}\;
  \caption{Compute enabler parameters}
  \label{algo:enabler_config}
\end{algorithm}

%% file: section/cost_table_example.tex
\begin{table*}[t]
  \setlength{\tabcolsep}{3pt}
  \footnotesize
  \centering
  \begin{tabular}{@{}|m{5.5cm}|m{8.2cm}|m{3.1cm}|@{}}
  \hline
    {\bf Configuration Constraint} & {\bf Cost} & {\bf Workload Predicate} \\
  \hline
  \hline
    {\tt autocommit!=0 \&\& flush\_log\_at\_trx\_commit==1} &\(2.6~\text{s}\), {\tt \{log\_write\_buf}$\rightarrow${\tt fil\_flush\}}, 17K syscalls, 100 I/O insts, $\ldots$ & \texttt{sql\_command==INSERT} \\
  \hline
    {\tt autocommit!=0 \&\& flush\_log\_at\_trx\_commit==2} & \(1.7~\text{s}\), {\tt \{log\_write\_buf\}}, 16.9K syscalls & \texttt{sql\_command==INSERT} \\
  \hline
    {\tt autocommit!=0 \&\& flush\_log\_at\_trx\_commit!=1 \&\& flush\_log\_at\_trx\_commit!=2} & \(1.2~\text{s}\), \texttt{\{\}}, 16.9K syscall & \texttt{sql\_command==INSERT} \\
  \hline
    {\tt autocommit==0} & \(0.6~\text{s}\), {\tt \{trx\_mark\_sql\_stat\_end\}}, 16.8K syscalls & \texttt{sql\_command==SELECT||$\ldots$} \\
  \hline
  \end{tabular}
  \caption{Example raw cost table \sys generates for \texttt{autocommit} parameter from symbolic execution of MySQL code in Figure~\ref{fig:autocommit_code}.}
	\label{tab:impact_example}
  \vspace{-0.15in}
\end{table*}

%% file: section/scale.tex
\section{Scaling \sys to Large Software}
\vspace{-0.05in}
In this section, we describe the challenges and our solutions for scaling \sys to 
large software. 

\vspace{-0.1in}
\subsection{Choice of Symbolic Execution Engine}
\vspace{-0.05in}
We initially build \sys on the KLEE~\cite{CadarKLEE2008} symbolic execution engine
because it is widely used and convenient to experiment with. However, while KLEE works 
well on moderate-sized programs, 
it cannot handle large programs like MySQL. KLEE models the environment
(POSIX runtime and libc) with simplified implementation. Large
programs use many libc or system calls that are unimplemented or
implemented partially/incorrectly, e.g., \texttt{fcntl}, \texttt{pread}, 
and socket. KLEE also does not support symbolic execution of multi-threaded
programs. We spent several months patching KLEE to fix the 
environment model and add multi-threading support. When we were finally
able to run MySQL with KLEE, it took 40 minutes to just pass initialization
even without symbolic data. 

We thus decided to switch to the \sse platform~\cite{ChipounovASPLOS2011}. 
\sse uses the real environment within an entire stack (OS kernel and libraries). 
Executing large software thus would encounter almost no compatibility 
issues. In addition, \sse uses QEMU and dynamic binary translation
to execute a target program. For instructions that access symbolic data, 
they are interpreted by the embedded KLEE engine; but instructions that
access concrete data are directly executed on host CPU.
Overall, while \sse's choice of using real environment in symbolic execution 
in general means slower analysis compared to using simplified models like KLEE, 
executing concrete instructions on host CPU offsets that slowness
and achieves significant speed-up. After migrating \sys to \sse and with some minor 
adjustments, we can start MySQL server within one minute. 

\vspace{-0.08in}
\subsection{Handle Complex Input Structure}
\label{sec:complex_input}
\vspace{-0.05in}
Since \badconfig is often only triggered by certain input, \sys makes 
input symbolic besides configuration. For small programs, the input type is typically 
simple, \emph{e.g.}, an integer, a string, which is easy to be made symbolic. 
But large programs' input can be complex structure 
like SQL queries. If we make the input variable symbolic, the program 
will be stuck in the input parsing code for a long time and the majority of the 
input generated is invalid. 
For example, we make input variable \texttt{char *packet} (32 bytes) 
in MySQL symbolic and execute it in \sse for 1 hour, which generates 
several hundred test cases, but none of which is a legal SQL query. Even
after adding some additional constraints, the result is similar.

This problem is not unique to our problem domain. Compiler testing~\cite{Csmith2011PLDI}
or fuzzing~\cite{oss_fuzz} also faces this challenge of how to generate
valid input to programs like C compiler or DBMS. We address this problem through 
a similar practice by introducing workload templates.
Instead of having the parser figure out a valid structure, we
pre-define a set of input templates that have valid structures. 
Then we parameterize the templates so that they are not fixed,
\emph{e.g.}, the query type, insertion value, the number of queries, etc.
In this way, we can make the workload template parameters symbolic.

\vspace{-0.08in}
\subsection{Reduce Profiling Overhead}
\vspace{-0.05in}
\label{sec:profile_overhead}
Profiling large programs can incur substantial overhead. We build \sys tracer 
using low-level signals emitted by \sse rather than intrusive
instrumentation. Nevertheless, symbolic execution is demanding for performance
analysis as the program runs much slower compared to native execution. Fortunately, \sys 
cares about the relative performance between different paths. We can still identify
\badconfig if the relative differences roughly match the native execution, 
which we find is true for most cases. 
\sys conducts differential analyses to capture performance anomalies. We describe three additional optimizations in \sys tracer.

First, the \sys tracer controls the start and end of its function
profiler. This is because if we enable the function profiler at the very
beginning, it can be overwhelmed by lots of irrelevant function calls. 
We add APIs in the tracer and will start the tracer when the target system 
finishes initialization and stop the tracer when the system enters the shutdown phase.

Second, the tracer avoids guest memory accesses and on-the-fly calculation. 
Accessing memory in an execution state goes through the emulated MMU in QEMU. 
\sys tracer only accesses and stores key information (most from
registers) about the call/return signals.  It defers the record matching,
call chain and latency calculation to path termination.

Third, \sys will disable state switching during latency tracking if necessary.
Since the function profiler calculates the execution time by subtracting the 
return signal timestamp from call signal timestamp, if \sse switches to 
execute another state in between, the recorded latency will include the 
state switching cost. This in general does not cause serious problems because
the costs occur in all states and roughly cancels out with our differential 
analysis. But in rare cases, the switching
costs can distort the results. When this happens, \sys will force 
\sse to disable state switching. 

\vspace{-0.08in}
\subsection{Path Explosion and Complex Constraints}
\vspace{-0.05in}
A common problem with symbolic execution is path explosion, especially 
when the symbolic value is used in library or system calls. In addition, 
some library calls with symbolic data yield complex constraints that make 
the symbolic execution engine spend a long time in solving the constraints.

\sys leverages a core feature in \sse, \emph{selective symbolic execution}~\cite{ChipounovASPLOS2011}, 
to address this problem. Selective symbolic execution allows transition between concrete 
and symbolic execution when crossing some execution boundary, \emph{e.g.}, a system call.
\sys uses the Strictly-Consistent Unit-Level Execution consistency model, which silently 
concretizes the symbolic value before entering the boundary and adds the concretized 
constraint to the symbolic value after exiting the boundary. This consistency 
model sacrifices completeness but it would not invalid the analysis result.
To improve completeness, we add some relaxation rules in \sys without causing 
functionality errors: 1) if the library call does not add side effect, such as \texttt{strlen}/\texttt{strcmp}, 
we make the return value symbolic and remove the concretized constraint; 2) 
if the library call has side effect but does not hurt the functionality, 
such as \texttt{printf}, we directly remove the concretized constraint.

One issue we encounter with the \sse silent 
concretization is that its \texttt{concretize} API will only concretize the 
symbolic variable. The symbolic variable can taint other variables (make them 
symbolic) when it is assigned to these variables, but these tainted 
variables are not concretized during silent concretization. Having these
tainted variables remain symbolic can add substantial overhead. We thus
add a new API in \sse, \texttt{concretizeAll}, that concretizes not only 
the given symbolic variable but also its tainted variables. We implement
this API by recording in each write operation a mapping from 
the symbolic expression to the target address in the memory object. Later when 
\texttt{concretizeAll} is called, we will look up the memory objects to find 
addresses that contain the same symbolic expression and also concretize them.

%% file: section/implement.tex
\vspace{-0.1in}
\section{Implementation}
\vspace{-0.05in}
\label{sec:implement}

We implement the major \sys components in C/C++. % with around 4000 lines of code. 
The \sys checker is implemented in Python. The \sys tracer is written as \sse plugins 
and leverages \sse's existing \texttt{LinuxMonitor} 
plugin and \texttt{FunctionMonitor} plugin to capture low-level signals. 
The \sys static analyzer is built on top of LLVM framework~\cite{Lattner2004CGO}.
The \sys trace analyzer is implemented as a standalone tool.

% One issue for the trace analyzer is that 
In function profiling, for efficiency, the tracer captures the addresses 
instead of names of invoked functions. This means the analyzer needs to 
resolve the addresses to names. % during critical path analysis. % The trace analyzer can use the symbol table of the target executable. But 
The problem is that the virtual address of the target program can change in each run. 
We address this issue by modifying the ELF loader of the \sse Linux kernel to expose the
\texttt{load\_bias}. Then the tracer will record the offset from
the \texttt{load\_bias}. % , i.e., \texttt{rawAddress - load\_bias}, 
The analyzer can then use the offsets to resolve the names.

% Support tracing of multi-threaded systems. To do so, we get the thread id of each 
% function call and keep different function call list for each thread. In 
% particular, we use the \texttt{getTid} function from the \texttt{LinuxMonitor} 
% plugin in S2E. 

%% file: section/evaluation.tex
\vspace{-0.08in}
\section{Evaluation}
\vspace{-0.05in}
\label{sec:evaluation}
We evaluate \sys to answer several key questions:
\begin{itemize}[noitemsep, leftmargin=*, topsep=0pt, partopsep=0pt]
\item How effective is \sys in detecting \badconfig?
\item Can \sys expose unknown \badconfig?
\item How useful is \sys's checker to the user?
\item What is the performance of \sys?
\end{itemize}

\begin{table}[t]
  \setlength{\tabcolsep}{3pt}
  \footnotesize
  \centering
  \begin{tabular}{@{}lllll@{\hspace{1pt}}l@{}l@{}}
  \toprule
  {\bf Software} & {\bf Desc.} & {\bf Arch.} & {\bf Version} & {\bf SLOC} & \rotatebox[origin=c]{20}{\bf Configs} & \rotatebox[origin=c]{20}{\bf Hook} \\
  \midrule
  MySQL & Database & Multi-thd & 5.5.59 & 1.2M & 330 & 197 \\
  Postgres & Database &  Multi-proc & 11.0 & 843K & 294 & 165 \\
  Apache & Web server & Multi-proc-thd & 2.4.38 & 199K & 172 & 158 \\
  Squid & Proxy server & Multi-thd & 4.1 & 178K & 327 & 96 \\
  \bottomrule
  \end{tabular}
  \caption{Evaluated software. Hook: SLOC of core \sys hooks.} % Multi-thd: multi-threaded. Multi-proc: multi-process.}
  \label{tab:eval_software}
  \vspace{-0.1in}
\end{table}
\input{section/eval_cases}

The experiments are conducted on servers with Dual Processor of Intel Xeon
E5-2630 (2.20GHz, 10 cores), 64 GB memory, 1 TB HDD running a Ubuntu 16.04.
Since \sse engine runs in QEMU, we create a guest image of Debian 9.2.1 x86\_64
with 4 GB memory for all the \sys tests.

\vspace{-0.08in}
\subsection{Target Systems}
\vspace{-0.05in}
We evaluate \sys on four popular and large (up to 1.2M SLOC) open-source
software (Table \ref{tab:eval_software}): MySQL, PostgreSQL, Apache, and Squid. 
\sys can successfully analyze large multi-threaded programs (MySQL and Squid) as well 
as multi-process (PostgreSQL, Apache) programs. 

The manual effort to use \sys on a target system is small, mainly required in 
two steps: (1) add configuration hooks (Section~\ref{sec:var_symbolic}); (2) supply 
input templates (Section~\ref{sec:complex_input}). The other steps in the workflow 
are automated. 

Table \ref{tab:eval_software} shows SLOC of the core hooks we add to the four
systems. The hook size varies across systems. MySQL hooks are largest in size
mainly because the system defines many (22) configuration types (\texttt{Sys\_var\_*}) so 
we need to add hook (about 7 SLOC) to each type. But the overall effort
for different systems is consistently small. The changes are typically contained
in a few places with other codes untouched. In addition,
most software rarely modifies the configuration data structure design, 
so the effort can carry through versions.

For (2), users typically already have some workload profiles. The effort needed is 
to parameterize and organize them into our format. In our experience 
with the four evaluated software, this process is straightforward and can be done 
in a few hours.

\vspace{-0.08in}
\subsection{Detecting Known Specious Config}
\vspace{-0.05in}
\label{sec:eval_known}
To evaluate the effectiveness of \sys we collect 17 \emph{real-world} 
\badconfig cases from the four systems. Table~\ref{tab:config_description}
lists the case descriptions. We collect them from ServerFault~\cite{serverfault}, 
dba~\cite{dba}, blog posts~\cite{percona}, and prior work~\cite{AttariyanOSDI2012}. 

For each case, \sys analyzes the related parameters and makes the config 
set and workload parameters symbolic and symbolically executes the system. 
Each test produces a trace. The \sys trace analyzer checks 
if any state performs significantly worse than other states. 
A case is detected when \sys explores at least one poor state in its trace \emph{and} the poor 
states enclose the problematic parameter value(s).

\begin{table}[t]
  \setcounter{magicrownumbers}{0}
  \setlength{\tabcolsep}{5pt}
  \footnotesize
  \centering
  \begin{tabular}{@{}llp{1cm}p{0.7cm}lp{1.2cm}p{1cm}p{0.7cm}@{}}
  \toprule
    {\bf Id.} & \rotatebox[origin=c]{90}{\bf Detect} & {\bf Explored States} & {\bf Poor States} & \rotatebox[origin=c]{90}{\parbox{0.9cm}{\bf Related\\Configs}} & {\bf Cost\hspace{0.2cm}Metrics} & {\bf Analysis Time} & {\bf Max Diff}\\
  \midrule
    c\rownumber & \cmark & 88 & 17 & 4 & Latency & \SI{6}{m}\SI{25}{s} & 14.5$\times$ \\
    c\rownumber & \cmark & 24 & 3 & 1 & Lat.\&Sync. & \SI{3}{m}\SI{13}{s} & 15.7$\times$ \\
    c\rownumber & \cmark & 224 & 88 & 5 & I/O & \SI{19}{m}\SI{41}{s} & 2.0$\times$ \\
    c\rownumber & \cmark & 787 & 100 & 2 & Latency & \SI{53}{m}\SI{50}{s} & 11.7$\times$ \\
    c\rownumber & \cmark & 494 & 44 & 3 & Latency & \SI{17}{m}\SI{56}{s} & 29.9$\times$ \\
    c\rownumber & \cmark & 891 & 12 & 5 & I/O &\SI{112}{m}\SI{24}{s} & 3.0$\times$ \\
    c\rownumber & \cmark & 89 & 7 & 2 & Lat.\&I/O & \SI{4}{m}\SI{6}{s} & 4.3$\times$ \\
    c\rownumber & \cmark & 195 & 8 & 3 & Latency & \SI{13}{m}\SI{8}{s} & 1.8$\times$ \\
    c\rownumber & \cmark & 110 & 2 & 3 & Lat.\&I/O & \SI{15}{m}\SI{20}{s} & 3.5$\times$ \\
    c\rownumber & \cmark & 231 & 13 & 7 & Latency & \SI{23}{m}\SI{30}{s} & 2.4$\times$ \\
    c\rownumber & \cmark & 61 & 9 & 2 & Latency & \SI{13}{m}\SI{17}{s} & 8.6$\times$ \\
    c\rownumber & \cmark & 34 & 4 & 2 & Latency & \SI{7}{m}\SI{15}{s} & 3.8$\times$ \\
    c\rownumber & \cmark & 50 & 5 & 3 & Latency & \SI{6}{m}\SI{10}{s} & 8.9$\times$ \\
    c\rownumber & \xmark & 112 & 0 & 2 & Latency & \SI{3}{m}\SI{42}{s} & 0.6$\times$ \\
    c\rownumber & \xmark & 23 & 0 & 3 & Latency & \SI{6}{m}\SI{12}{s} & 0.2$\times$ \\
    c\rownumber & \cmark & 81 & 1 & 0 & Latency & \SI{433}{m}\SI{32}{s} & 4.3$\times$ \\
    c\rownumber & \cmark & 3 & 1 & 0 & I/O & \SI{1}{m}\SI{32}{s} & 2.0$\times$ \\
  \bottomrule
  \end{tabular}
  \caption{\sys detection result. Poor states are what \sys considers as potential \badconfig, so 0 means undetected.}
  \label{tab:main_result}
\end{table}

In total, \sys detects 15 of the 17 cases. Table~\ref{tab:main_result} shows the detailed result. 
For each case, Table~\ref{tab:main_result} lists the total states \sys explored, 
poor states, related configs, and maximum cost metric differences. 
The explored states include forks from related configurations and the symbolic workload parameters. 
In most cases, the \badconfig requires specific related settings 
to expose the issue. The high success rate of \sys comes from 
its in-vivo multi-path profiling, dependency analysis, and differential 
performance analysis. 

Another aspect to interpret the high success rate is that the 17 cases we collect
admittedly have a selection bias---all cases cause severe performance impact. 
This is reflected in the max diff column of the cases \sys detects. If a
misconfiguration only introduces mild performance issue,
\sys may miss it due to the noises in symbolic execution. 
However, \sys's goal is to exactly target \badconfig that has severe 
performance impact, rather than suboptimal configurations.

\sys misses two Apache cases, c14 and c15. Triggering them requires
enabling the HTTP KeepAlive feature in the workload. In our Apache workload 
templates, this feature is not part of the workload parameters 
and is disabled by default.

We describe two representative cases. MySQL c1 is the running example in the paper.
\sys identifies four related parameters like \texttt{flush\_at\_trx\_commit} 
that together affect the performance. \sys explores 88 states 
in total, 4 of which are identified as poor. The configuration constraints 
of the four poor states describe the combination conditions for the 5 parameters 
to incur significant cost.

In c6, \texttt{innodb\_log\_buffer\_size} controls the size of the log buffer. 
\sys identifies that the \texttt{innodb\_flush\_method}, \texttt{autocommit}, 
\texttt{flush\_at\_trx\_commit} and two other parameters are related. 
Interestingly, in this case, \sys determines the latency metric difference is
not significant, but the I/O logical cost metric is. Specifically, \sys explores 
almost 100 different queries, and finds that in states with queries involving 
large row changes and a relatively small buffer size, 
the I/O metric---\texttt{pwrite} operations---is much larger than other states.

\vspace{-0.08in}
\subsection{Comparison with Testing}
\vspace{-0.08in}
\label{sec:testing}
We evaluate the 17 cases with testing as well. We use popular benchmark tools 
\texttt{sysbench} and \texttt{ab}. 
For each case, we set the target parameter and related parameters with concrete 
values from one of the poor states \sys discovers.
We then enumerate the standard workloads in the benchmark to test the 
software with the configurations. The performance numbers produced by 
testing are absolute, which are difficult to judge. 
We use configurations from the good states and collect performance with them as a baseline. 
If the performance difference ratio exceeds 100\% (the same threshold used by \sys), 
we consider the case detected. In total, testing 
detects 10 cases, with a median time of 25 minutes.

\sys is not meant to replace configuration performance testing. Given enough 
time and resources, exhaustive performance testing 
can in theory expose all cases. But in practice, users cannot afford exhaustive 
testing. \sys systematically explores program states while 
avoids the redundancy in exhaustive testing (Section~\ref{sec:se_background}).
Even though in some cases, as shown in Table~\ref{tab:main_result}, the 
\sys analysis time is relatively long, \sys is exploring the performance 
effects thoroughly, including the combination effect with other 
parameters and input. Therefore, the performance impact models \sys derives 
are complete. Once the exploration is done, the outcome can be reused many 
times while testing needs to be done repeatedly. 

Another challenge with testing is how to judge whether the absolute performance results of 
a configuration are good or not. Our experiment above assumes the existence
of a good configuration, which users may not have. \sys, in comparison, 
conducts in-vivo, multi-path analysis, so it naturally has baselines to 
compare with. The analysis enables \sys to collect deeper logical metrics, 
which can reveal performance issues that end-to-end metrics may not find.

\begin{figure*}[t]
  \centering
  \begin{minipage}[t]{0.33\textwidth}
  \includegraphics[width=\textwidth]{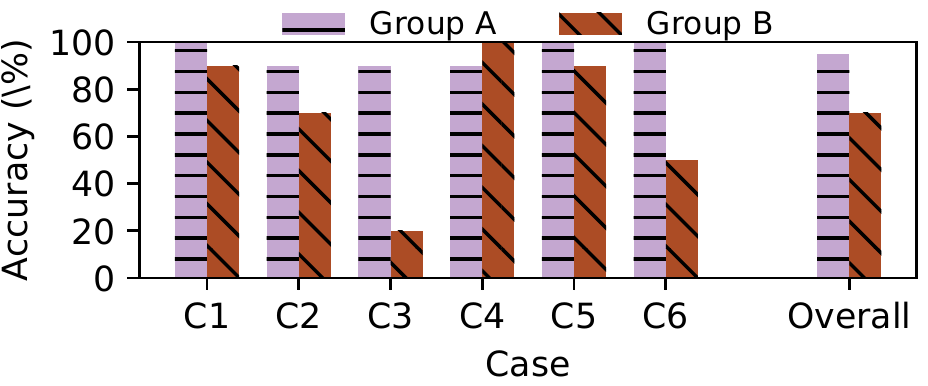}
  \caption{Overall accuracy of judgment in the user study.}
  \label{fig:user_study_accuracy}
  \end{minipage}
  \quad
  \begin{minipage}[t]{0.31\textwidth}
    \includegraphics[width=\textwidth]{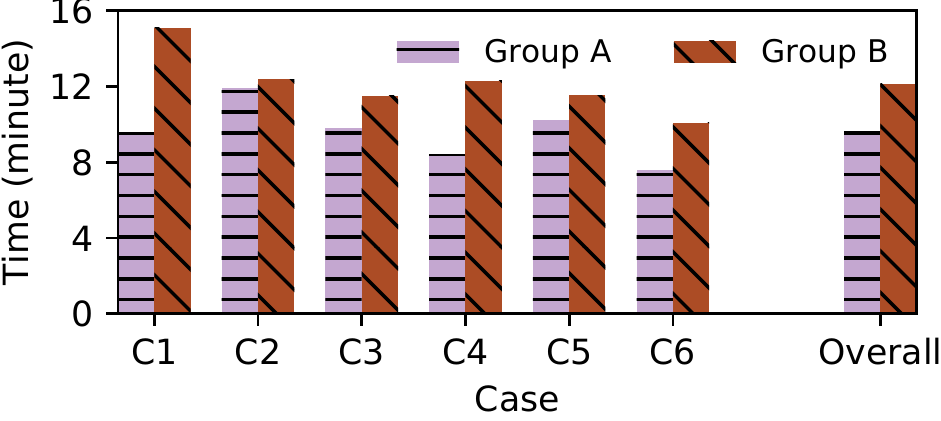}
    \caption{Average decision time in the user study.}
    \label{fig:user_study_time}
  \end{minipage}
  \quad
  \begin{minipage}[t]{0.3\textwidth}
    \includegraphics[width=\textwidth]{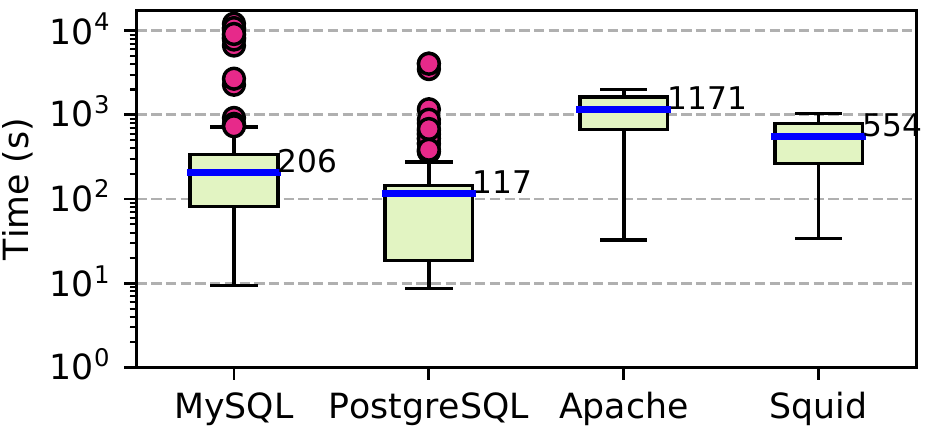}
    \caption{\sys analysis times for the configs in the four software.}
    \label{fig:coverage_analysis_time}
  \end{minipage}
  \vspace{-0.15in}
\end{figure*}

\vspace{-0.08in}
\subsection{Exposing Unknown Specious Config}
\vspace{-0.05in}
\label{sec:eval_unknown}

Besides detecting know \badconfig, we evaluate whether \sys can expose
unknown \badconfig. We first apply \sys to derive performance models 
for all parameters if possible (Section~\ref{sec:coverage}).
We then analyze the results for parameters not in the known case 
dataset (Section~\ref{sec:eval_known}). We manually check 
(1) if some parameter's default or suggested value is in a poor state; 
(2) if a poor state of a parameter contains related parameters that are undocumented. 
The manual inspection involves checking the \sys output, the descriptions in
the official documentation and tuning guide, and running tests to
confirm, which takes significant time. We only carefully inspect a
subset of the results. 

The four systems are very mature and maintain high-quality 
documentations, so it is not easy to find many errors in them. Indeed, a significant 
portion of the poor states we examined turns out to be already documented. 
But still we have identified 9 parameters that have potential 
bad performance effect and the documentation is 
incomplete or incorrect. 

\begin{table}
  \setlength{\tabcolsep}{3pt}
  \footnotesize
  \centering
  \begin{tabular}{@{}llp{4.8cm}@{}}
  \toprule
    \textbf{Sys} & \textbf{Configuration} & \textbf{Performance Impact} \\
  \midrule
    \multirow{2}{*}{Postgres} & \texttt{vacuum\_cost} & Default value \SI{20}{ms} is significantly worse \\
    & \texttt{\_delay} & than low values for write workload. \\
    Postgres & \texttt{archive\_timeout} & Small values cause performance penalties.\\
    Postgres & \texttt{random\_page\_cost} & Values larger than 1.2 (default 4.0) cause bad perf on SSD for join queries.\\
    Postgres & \texttt{log\_statement} & Setting \texttt{mod} causes bad perf. for write workload 
    when \texttt{synchronous\_commit} is \texttt{off}.\\
    \multirow{2}{*}{Postgres} & \texttt{parallel\_leader} & Enabling it can cause select join query\\
    & \texttt{\_participation} & to be slow if \texttt{random\_page\_cost} is high.\\
    \multirow{2}{*}{MySQL} & \texttt{optimizer\_search} & Default value would cause bad performance \\
    & \texttt{\_depth} & for join queries\\
    MySQL & \texttt{concurrent\_insert} &  Enable concurrent\_insert would cause bad performance for read workload \\
    Squid & \texttt{ipcache\_size} & The default value is relatively small and may cause performance reduction\\
    Squid & \texttt{cache\_log} & Enable cache\_log with higher debug\_option would cause extra I/O\\
  \bottomrule
  \end{tabular}
  \caption{Unknown perf. effect of 9 parameters \sys identifies.}
  \label{tab:unknown_cases}
\end{table}

Table~\ref{tab:unknown_cases} lists the cases. For example, 
our analysis of \texttt{vacuum\_cost\_delay}
shows that a higher value can incur large cost
for write-intensive workloads. But the default value is 
\SI{20}{ms}. 
For \texttt{log\_statement}, \sys discovers multiple poor states. The official
document does not mention its performance impact. 
Our analysis 
revels that setting it to \texttt{mod} causes performance 
issues for write workloads when \texttt{synchronous\_commit} is 
\texttt{off}. \sys finds some unexpected parameter combination 
that leads to bad performance, e.g., \texttt{parallel\_leader\_participation}
and \texttt{random\_page\_cost}.

We reported our findings to the developers. Seven reports are confirmed. 
Three lead to documentation or Wiki fixes. For some confirmed cases, developers do not fix them because they 
assume users should know the performance implications or such performance 
description should not be put in the reference manual (e.g., \emph{``There are a lot of 
interactions between settings, and mentioning all of them would be impossible''}).

\vspace{-0.08in}
\subsection{User Study on \sys Checker}
\vspace{-0.05in}
To understand whether \sys checker can help users catch \badconfig, we conduct
a controlled user study with 20 programmers (no authors are included). Fourteen
of them are undergraduate students who have taken the database class. Six are
graduate students. They all have decent experience with database and Unix
tools. We further give a tutorial of the MySQL and PostgreSQL background, the
descriptions of the common configuration parameters, and available benchmark
tools they can use.  We acknowledge that the representativeness of our study
participants may not be perfect.

We use 6 target parameters from MySQL and PostgreSQL. 
For each parameter, we prepare two versions of configuration files. 
In one version (bad), the parameter is set with the poor value and the
related parameters are also set appropriately that would cause bad performance
impact under a workload. In another version (good), we set the
target parameter to a good value, or we change the related
parameter values, or we tell users the production workloads are limited
to certain types (e.g., read-intensive). So in total, we have 12 cases.

Each participant is given 6 configuration files. They need to 
make a judgment regarding whether the configuration file would cause 
potential performance issue. Since a configuration file contains 
many parameters, we explicitly tell users the set of parameters 
they can focus on, which disadvantages \sys because users 
in practice do not have this luxury.

The participants are randomly assigned into two groups: \emph{group A} (w/ \sys 
checker help) and \emph{group B} (w/o checker help). Users in group B can run 
any tools to help them make the decision. We also tell 
group A users that they do not have to trust the checker output and are free to 
run other tools. 

Figure~\ref{fig:user_study_accuracy} shows the accuracy of user study result
for each group. Overall, programmers w/o \sys checker's help have 30\% 
misjudgment rate while programmers with \sys checker's help only 
have 5\% misjudging rate. Figure~\ref{fig:user_study_time} shows the time 
for making a judgment. On average, participants took 20.7\% less time 
(9.6 minutes versus 12.1 minutes) to make a judgment when they were provided 
with \sys checker. 

\vspace{-0.08in}
\subsection{Coverage of Analyzed Configs}
\vspace{-0.05in}
\label{sec:coverage}
\begin{table}
  \footnotesize
  \setlength{\tabcolsep}{5pt}
  \centering
  \aboverulesep=0ex
  \belowrulesep=0ex
  \begin{tabular}{@{}llll|l@{}}
  \toprule
  MySQL & PostgreSQL & Apache & Squid & Total \\
  \midrule
  169 (51.2\%) & 210 (71.4\%) & 51 (29.6\%) & 176 (53.3\%) & 606 (53.9\%) \\
  \bottomrule
  \end{tabular}
  \caption{Number of configs \sys derives performance models for. 
  The number in parentheses is the percentage of total configs.}
  \label{tab:config_coverage}
\end{table}

We conduct a coverage test of \sys by applying \sys on the four software and 
try to derive performance models for as many parameters as possible. 
We manually filter the parameters that are not related to performance based on the parameter description
(\emph{e.g.}, \texttt{listen\_addresses}). Table~\ref{tab:config_coverage} shows the result.
\sys successfully derives models for a total of 606 parameters. 
The average ratio of analyzed parameters over the total number of parameters 
for software is 53.9\%. The average number of states explored in these
generated models is 23. Apache and Squid have a relatively small number of
parameters analyzed. This is because the configuration program variables 
in the two systems are set via complex function pointers and spread in 
different modules, which make it challenging to write hooks to enumerate 
all of them (Section~\ref{sec:var_symbolic}). For parameters that \sys did 
not generate impact models, one reason is that they are used in code for
special environment. Another reason is that the data type of some parameter is
too complex (\emph{e.g.}, timezone) to make symbolic.

\vspace{-0.08in}
\subsection{Accuracy of \sys Profiling}
\vspace{-0.05in}

\begin{table}
  \footnotesize
  \setlength{\tabcolsep}{2pt}
  \centering
  \aboverulesep=0ex
  \belowrulesep=0ex
  \begin{tabular}{@{}c@{\hspace{8pt}}rrcrrcrrrcrrr@{}}
  \toprule
    & \multicolumn{2}{c}{parA} & \phantom{a} & \multicolumn{2}{c}{parB} & \phantom{a} & \multicolumn{3}{c}{parC} & \phantom{a} & \multicolumn{3}{c}{parD} \\
  \cmidrule{2-3} \cmidrule{5-6} \cmidrule{8-10} \cmidrule{12-14}
    & =0 & =1 & & =0 & =1 & & =0 & =1 & =2 & & =0 & =1 & =2 \\
  \midrule
  {\bf Violet} & 12.0 & 23.0 & & 9.81 & 10.19 & & 9.05 & 10.92 & 10.74 & & 4.68 & 4.77 & 5.27 \\
    {\bf \sse} & 10.8 & 21.0 & & 7.67 & 8.94 & & 6.24 & 7.77 & 7.92 & & 3.57 & 3.91 & 4.59 \\
    {\bf Native} & 0.7 & 1.2 & & 0.55 & 0.77 & & 0.45 & 0.63 & 0.67 & & 0.07 & 0.07 & 0.08 \\
  \bottomrule
  \end{tabular}
  \caption{Absolute latency (ms) for four parameters' different settings w/ \sys, 
  vanilla \sse and native execution. parA: \texttt{autcommit}, parB: \texttt{synchronous\_commit}, 
  parC: \texttt{archive\_mode}, parD: \texttt{HostNameLookup}.}
  \label{tab:accuracy}
\end{table}

Since symbolic execution can introduce significant overhead, it seems that the 
latency traced by the symbolic engine cannot be very accurate. However, 
we observe that while the absolute latency under symbolic execution is indeed much 
larger than native execution, the comparative results between different paths are 
usually similar. We add a micro-benchmark experiment to test the latency measurement 
from \sys, vanilla \sse and native mode. Table ~\ref{tab:accuracy} shows the result 
from four representative parameters. Take parA as an example. The latency results 
from \sys and \sse are much later than native result. But the ration of setting 1 to 
setting 0 is similar: 1.92$\times$ for \sys, 1.94$\times$ for \sse, and 1.71$\times$ 
for native execution.

\vspace{-0.08in}
\subsection{False Positives}
\vspace{-0.05in}
The \sys differential performance analysis in general can absorb the performance 
noises in symbolic execution. But we observe some false positives in the
\sys performance analysis output. For example, \sse somehow has a delay 
in emitting the return signal of some system call functions like \texttt{gettimeofday},
which causes \sys to record inaccurate latency. These false positives are 
relatively easy to suppress by discounting the cost of the noisy instructions.

We manually inspect the performance models of 10 random parameters that \sys analyzes in the coverage experiment. 
We check the accuracy of the reported bad states by verifying them with \texttt{sysbench}. The 
false positive rate is 6.4\%.

\vspace{-0.1in}
\subsection{Performance}
\vspace{-0.05in}

We measure the \sys analysis time for the 471 parameters in the coverage 
experiment (Section~\ref{sec:coverage}). Figure~\ref{fig:coverage_analysis_time} 
shows the result in boxplots. The median analysis times are 
\SI{206}{s} (MySQL), \SI{117}{s} (PostgreSQL), \SI{1171}{s} (Apache), 
and \SI{554}{s} (Squid). On average, the log analyzer time is 68s. 
As explained in Section~\ref{sec:testing}, even though
for some parameters the analysis time is relatively long, the benefit is that
\sys derives a thorough performance model for different settings of the target 
parameter and the combined effect with other parameters and input.
The outcome can be re-used many times by the \sys checker. With the 
performance models, the checker time is fast. On average the checking 
only takes 15.7 seconds.

\begin{figure}[t]
  \centering
  \includegraphics[width=3.25in]{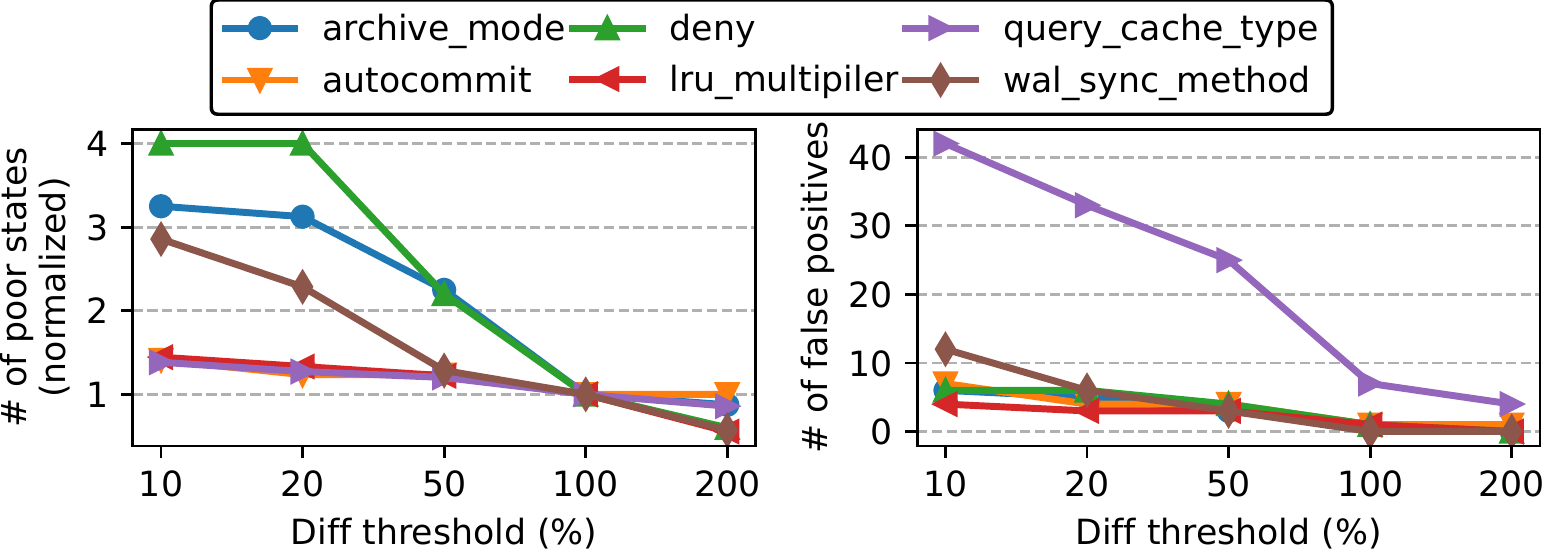}
  \caption{Sensitivity of the performance diff threshold (default 100\%). 
   For readability, the number of poor states is normalized by values under
   the default threshold.}
  \label{fig:sensitivity}
\end{figure}

\vspace{-0.08in}
\subsection{Sensitivity Analysis}
\vspace{-0.05in}
\sys uses a differential threshold (default 100\%) to detect the suspicious state from the trace 
log (Section~\ref{sec:trace_analysis}). We evaluate the sensitivity of 
this threshold. We measure how many poor state pairs \sys reports 
when analyzing a parameter under threshold $t$. Then for each poor state pair \sys 
reports, we run benchmarks on the native machine to check whether it is false 
positive (performance difference is $\geq t\%$). 

Figure~\ref{fig:sensitivity} shows the result for six representative parameters. We can 
see that if the threshold is set to a relatively lower value, the 
number of detected \badconfig can dramatically increase, but at cost of higher 
false positives. 

%% file: section/eval_cases.tex
\begin{table*}[!tbp]
  \setlength{\tabcolsep}{1ex}
  \setcounter{magicrownumbers}{0}
  \footnotesize
  \centering
  \begin{tabular}{@{}lm{1.5cm}>{\ttfamily}m{3.5cm}m{1.5cm}m{9.5cm}@{}}
   \toprule
    {\bf Id.} & {\bf Application} & {\bf Configuration Name} & {\bf Data Type} & {\bf Description}\\
   \midrule
    c\rownumber & MySQL & autocommit & Boolean & Determine whether all changes take effect immediately \\
  %  c2 & MySQL & sql\_log\_bin & Boolean & Control whether logging to the binary log is enabled \\
    c\rownumber & MySQL & query\_cache\_wlock\_invalidate & Boolean & Disable the query cache when after WRITE lock statement \\
    c\rownumber & MySQL & general\_log & Boolean & Enable MySQL general log query \\
    c\rownumber & MySQL & query\_cache\_type & Enumeration & Method used for controlling the query cache type \\
    c\rownumber & MySQL & sync\_binlog & Integer & Controls how often the MySQL server synchronizes binary log to disk \\
    c\rownumber & MySQL & innodb\_log\_buffer\_size & Integer & Set the size of the buffer for transactions that have not been committed yet \\ 
  \hline
  \hline
  %  c\rownumber & PostgreSQL & synchronous\_commit & Boolean & Specifies whether transaction commit will wait for WAL records to be written to disk
  %   before the command returns a ``success'' indication to the client \\
  %  c\rownumber & PostgreSQL & wal\_writer\_delay  & Integer & Specifies how often the WAL writer flushes WAL \\
    c\rownumber & PostgreSQL & wal\_sync\_method & Enumeration & Method used for forcing WAL updates out to disk \\
  %  c\rownumber & PostgreSQL & log\_statement & Enumeration & Control when to log the statement \\
    c\rownumber & PostgreSQL & archive\_mode & Enumeration & Force the server to swtich to a new WAL periodically and archive old WAL segments\\
    c\rownumber & PostgreSQL & max\_wal\_size & Integer & Maximum number of log file segments between automatic WAL checkpoints \\
    c\rownumber & PostgreSQL & checkpoint\_completion\_target & Float & Set a fraction of total time between checkpoints interval \\
    c\rownumber & PostgreSQL & bgwriter\_lru\_multiplier & Float & Set estimate of the number of buffers for the next background writing\\
%    c\rownumber & PostgreSQL & random\_page\_cost & Float & Sets the estimate of the cost of a non-sequentially-fetched disk page \\
  \hline
  \hline
    c\rownumber & Apache & HostNamelookup & Enumeration & Enables DNS lookups to log the host names of clients sending requests \\
    c\rownumber & Apache & Deny/Domain & Enum/String & Restrict access to the server based on hostname, IP address, or env variables \\
    c\rownumber & Apache & MaxKeepAliveRequests & Integer & Limits the number of requests allowed per connection \\
    c\rownumber & Apache & KeepAliveTimeOut & Integer & Seconds Apache will wait for a subsequent request before closing the connection \\
  \hline
  \hline
    c\rownumber & Squid & cache & String & Requests denied by this directive will not be stored in the cache\\
    c\rownumber & Squid & Buffered\_logs & Integer & Whether to write access\_log records ASAP or accumulate them in larger chunks\\
  \bottomrule
  \end{tabular}
  \caption{Description of 17 known \badconfig cases we collect in the four evaluated software.}
  \label{tab:config_description}
  \vspace{-0.1in}
\end{table*}

% \begin{table*}[!tbp]
%   \setlength{\tabcolsep}{1ex}
%   \setcounter{magicrownumbers}{0}
%   \footnotesize
%   \centering
%   \begin{tabular}{@{}m{3.5cm}m{1cm}m{1cm}m{1cm}m{1cm}m{1cm}m{1cm}m{1cm}m{1cm}m{1cm}m{1cm}@{}}
%    \toprule
%     \multirow{2}{2cm} {\bf Configuration Name} & \multicolumn{5}{c}{\bf Detect Cases} & \multicolumn{5}{c}{\bf False Positive}\\
%     & T=0.1 & T=0.2 & T=0.5 & T=1 & T=2  & T=0.1 & T=0.2 & T=0.5 & T=1 & T=2\\ 
%    \midrule
%     autocommit & 24 & 21 & 21 & 17 & 17 & 7 & 4 & 4 & 1 & 1 \\
%     query\_cache\_type& 137 & 126  & 119 & 99 & 85 & 42 & 33 & 25 &7  &4 \\
%     wal\_sync\_method& 20 & 16 & 9 & 7 & 4 & 12 & 6 & 3 &0 &0 \\
%     archive\_mode & 26 & 25 & 18 & 8 & 7 & 6 & 5 & 3 & 1& 0 \\
%     bgwriter\_lru\_multipiler & 13 & 12 & 11 & 9  & 5 & 4 & 3 &3 & 1&0 \\
%     HostNameLookup & 6 & 6 & 6 & 4  & 4 & 1 & 1 & 1 & 0& 0\\
%     deny & 20 & 20 & 11 & 5 & 3 & 6 & 6 & 4 & 1 & 0 \\
%   \bottomrule
%   \end{tabular}
%   \caption{The sentivity of threshold}
%   \label{tab:sentivity}
%   \vspace{-0.1in}
% \end{table*}

%% file: section/discuss.tex
\vspace{-0.08in}
\section{Limitations}
\vspace{-0.08in}
\label{sec:discussion}
\sys has several limitations that we plan to address in future work. First,
\sys explores the configuration under normal conditions.
% \sys uses normal symbolic execution environment that \sys uses to explore
Some \badconfig may be only used in error handling. % For example,
% \texttt{minimal\_backoff\_time} in Postfix controls the time to back off from a
% dead destination. 
Exploring their effect requires specific faults. One solution is to combine symbolic execution with fault injection. Another 
potential solution is to use under-constrained symbolic execution~\cite{UCSE2015Sec}.
Second, our handling of floating point type parameters is imperfect due to 
limited support in existing symbolic execution engines. We currently explores
float parameters by choosing from a set of concrete floating-point values
in the valid value range. Third, we use concrete (the host) hardware 
in the symbolic execution, which may not capture \badconfig that is only visible 
in specific hardware. We rely on logical cost metrics to surface such issues. 
Lastly, \sys does not work on distributed systems. %An improvement could be using abstract hardware models.

% (e.g., SSD but not HDD)
% some parameters' performance effect depends
% on the 

% For example, \texttt{minimal\_backoff\_time} in Postfix is the
% minimal amount of time a message will not be looked at, and the minimal amount
% of time to stay away from a dead destination. The default value of this
% parameter before postfix 2.4 is \SI{1000}{\second}, which can cause slow
% message delivery on errors. The default is changed to \SI{300}{\second} in
% recent versions. 

%% file: section/related.tex
\vspace{-0.08in}
\section{Related Work}
\label{sec:related}
\vspace{-0.1in}
\boldhdr{Misconfiguration detection and diagnosis}
A wide body of work has been done to detect and troubleshoot misconfiguration~\cite{
Wang2003LISA,Whitaker2004OSDI,Wang2004OSDI,FeamsterNSDI2005,Su2007SOSP,DasSecurity2010,FeamsterNSDI2005,
BauerSACMAT2008,Attariyan2008,YuanATC2011,Attariyan2010,ZhangASPLOS2014}. For example,
ConfAid~\cite{Attariyan2010} uses dynamic taint tracking to locate configuration errors 
that lead to failures; Strider~\cite{Wang2003LISA} and PeerPressure~\cite{Wang2004OSDI} take 
statistical approaches to identify misconfiguration; EnCore~\cite{ZhangASPLOS2014} 
enhances statistical learning with environment information to detect misconfiguration. 

These solutions mainly target illegal configuration and have limited effects on 
\badconfig. 
X-ray~\cite{AttariyanOSDI2012} targets performance-related misconfiguration. 
Our work is inspired by X-ray and is complementary to it. X-ray is a diagnosis tool 
and uses deterministic record and replay of a specific program execution. 
\sys focuses on detecting \badconfig beforehand. \sys uses symbolic execution 
to explore the performance effect in multiple execution paths. \sys is 
more suitable for performance tuning/bug finding, whereas X-ray is better 
at diagnosing misconfiguration that has occurred.

LearnConf~\cite{LearnConfEuroSys2020} is recently proposed to detect performance misconfiguration
using static analysis. LearnConf summarizes common code patterns of 
performance misconfiguration and uses simple formulas to approximate the performance 
effect, \emph{e.g.}, linear relationship. It uses static analysis to identify these 
patterns and derive parameters to the formulas. 
The solution is simpler compared to \sys. But its completeness is limited because
obtaining comprehensive code patterns is hard. Moreover, the
performance effect is often quite complex, which cannot be
accurately captured by simple formulas. Static analysis also 
suffers from well-known inaccuracies for large software. \sys explores a
configuration's influence in the code without requiring or being limited by
common patterns; it analyzes the performance effect by executing the code.
Additionally, Violet explores the performance impact of input and a large set
of related configurations together.

\sboldhdr{0.04in}{Performance tuning of configuration}
There is a wealth of literature on automatic performance tuning, e.g., 
\cite{SmartConf2018,CDBTune2019,Herodotou11starfish,YeSIGMETRICS2003,XiWWW2004,
OsogamiSIGMETRIC2006,BestConfig2017}. They work basically by devising an approximate 
function between configuration values and the performance metrics measured through testing. 
While tunable parameters are common \badconfig, performance tuning and detecting \badconfig are two 
directions. The former searches for settings that yield the best 
performance, while the latter identifies settings that lead to extremely poor performance.
\sys takes an analytical approach to derive configuration performance impact
model from the code, instead of exhaustive testing. The result from our in-vivo, multi-path
analysis is also less susceptible to noises and enables extrapolation
to different contexts.

\sboldhdr{0.04in}{System resilience to misconfiguration} 
ConfErr~\cite{KellerDSN2008} uses a human error model to 
inject misconfiguration. SPEX~\cite{Xu2013SOSP} uses static analysis to 
extract configuration constraints and generates misconfiguration by 
violating these constraints. 
The injected misconfigurations are illegal values that can trigger explicit errors 
like crash. Specious configuration typically does not cause explicit errors.

\sboldhdr{0.04in}{Configuration languages}
Better configuration languages can help avoid misconfiguration. Several works make such 
efforts~\cite{DeTrevilleHOTOS2005,LooSIGCOMM2005,BeckettSIGCOMM2016,EnckATC2007,ChenCONEXT2010,schpbach2011a,HuangEuroSys2015}. 
PRESTO~\cite{EnckATC2007} proposes a template language to generate device-native 
configuration. 
ConfValley~\cite{HuangEuroSys2015}, proposes a declarative validation language 
for generic software configuration. 
These new designs do not prevent \badconfig from being introduced.

\sboldhdr{0.04in}{Symbolic execution in performance analysis} Symbolic execution~\cite{KingCACM1976,CadarKLEE2008} 
is typically used for finding functional bugs. \sse~\cite{ChipounovASPLOS2011} is 
the first to explore performance analysis in symbolic execution as one use case 
to demonstrate the generality of its platform. The \sys tracer leverages the advances 
made by \sse, particularly its low-level signals, to build our custom profiling 
methods (Section~\ref{sec:profile_path}). Our tracer also addresses several 
unique challenges to reduce the performance analysis overhead (Section~\ref{sec:profile_overhead}).
Bolt~\cite{PerfContractsNSDI2019} extracts performance contracts of Network Function code
with symbolic execution. \sys targets general-purpose software and analyzes 
performance effect of system configuration.

%% file: section/conclusion.tex
\vspace{-0.08in}
\section{Conclusion}
\label{sec:conclusion}
\vspace{-0.05in}
Specious configuration is a common and challenging problem for production
systems. % Its subtle nature makes it difficult. % to detect the misconfiguration
% without careful reasoning about performance and considerations of the exact conditions depending on various factors. 
We propose an analytical approach to tackle this problem and 
present a toolchain called \sys. \sys uses symbolic execution 
and program analysis to systematically reason about the performance effect 
of configuration from code. %, as well as the conditions to trigger the performance issue. 
The derived configuration performance impact model can be used for subsequent 
detections of \badconfig. We successfully apply \sys on four large system software 
and detect 15 out of 17 real-world \badconfig cases. \sys exposes 9 unknown \badconfig, 
7 of which are confirmed by developers.

%% file: section/ack.tex
\vspace{-0.1in}
\section*{Acknowledgments}
\vspace{-0.08in}
\label{sec:ack}
We would like to thank our shepherd, Jason Flinn, and the anonymous OSDI reviewers 
for their thoughtful comments. We appreciate the 
discussion and suggestions from Xi Wang. We thank Varun Radhakrishnan and Justin Shafer 
for their contributions to the Violet tool and study cases. We thank our 
user-study participants and the open-source developers who responded to 
our requests. We also thank the \sse authors, especially Vitaly Chipounov for 
maintaining the \sse platform and answering our questions. We thank Chunqiang Tang 
for the prior collaboration that provided early motivation for this work. 
This work is supported by the NSF CRII grant CNS-1755737 and partly by 
NSF grant CNS-1910133.

%% file: main.bbl
\begin{thebibliography}{10}

\bibitem{aws_feb_2017}
{Amazon AWS S3} outage for several hours on {February} 28th, 2017.
\newblock \url{https://aws.amazon.com/message/41926}.

\bibitem{aws_april_2011}
{Amazon EC2} and {RDS} service disruption on {April} 21st, 2011.
\newblock \url{http://aws.amazon.com/message/65648}.

\bibitem{aws_october_2012}
{AWS} service outage on {October} 22nd, 2012.
\newblock \url{https://aws.amazon.com/message/680342}.

\bibitem{dba}
Database administrators.
\newblock \url{https://dba.stackexchange.com}.

\bibitem{facebook_sept_2010}
{Facebook} global outage for 2.5 hours on {September} 23rd, 2010.
\newblock
  \url{https://www.facebook.com/notes/facebook-engineering/more-details-on-todays-outage/431441338919}.

\bibitem{google_april_2013}
{Google API} infrastructure outage on {April} 30th, 2013.
\newblock
  \url{http://googledevelopers.blogspot.com/2013/05/google-api-infrastructure-outage_3.html}.

\bibitem{google_cloud_april_2016}
Google compute engine incident \#16007.
\newblock
  \url{https://status.cloud.google.com/incident/compute/16007?post-mortem}.

\bibitem{google_jan_2014}
{Google} service outage on {January} 24th, 2014.
\newblock
  \url{http://googleblog.blogspot.com/2014/01/todays-outage-for-several-google.html}.

\bibitem{azure_december_2012}
{Microsoft Azure} storage disruption in {US} south on {December} 28th, 2012.
\newblock
  \url{http://blogs.msdn.com/b/windowsazure/archive/2013/01/16/details-of-the-december-28th-2012-windows-azure-storage-disruption-in-us-south.aspx}.

\bibitem{azure_february_2013}
{Microsoft Azure} storage disruption on {February} 22nd, 2013.
\newblock
  \url{http://blogs.msdn.com/b/windowsazure/archive/2013/03/01/details-of-the-february-22nd-2013-windows-azure-storage-disruption.aspx}.

\bibitem{oss_fuzz}
Oss-fuzz: Continuous fuzzing for open source software.
\newblock \url{https://github.com/google/oss-fuzz}.

\bibitem{percona}
Percona blogs.
\newblock \url{https://www.percona.com/blog}.

\bibitem{rds_mysql}
{RDS MySQL} insights: Top query "commit".
\newblock
  \url{https://serverfault.com/questions/1029595/rds-mysql-insights-top-query-commit}.

\bibitem{serverfault}
Serverfault.
\newblock \url{https://serverfault.com}.

\bibitem{slow_insert_update}
Slow {InnoDB} insert/update.
\newblock
  \url{https://www.serveradminblog.com/2014/01/slow-innodb-insertupdate/}.

\bibitem{sysbench}
Sysbench.
\newblock \url{https://github.com/akopytov/sysbench}.

\bibitem{cisco_august_2017}
Cisco loses customer data in {Meraki} cloud muckup due to misconfiguration.
\newblock
  \url{https://www.theregister.co.uk/2017/08/06/cisco_meraki_data_loss}, Aug
  6th, 2017.

\bibitem{aws_december_2012}
Amazon.
\newblock {AWS} service outage on {December} 24th, 2012.
\newblock \url{http://aws.amazon.com/message/680587}.

\bibitem{AttariyanOSDI2012}
M.~Attariyan, M.~Chow, and J.~Flinn.
\newblock X-ray: Automating root-cause diagnosis of performance anomalies in
  production software.
\newblock In {\em Proceedings of the 10th USENIX Conference on Operating
  Systems Design and Implementation}, OSDI'12, pages 307--320, 2012.

\bibitem{Attariyan2008}
M.~Attariyan and J.~Flinn.
\newblock Using causality to diagnose configuration bugs.
\newblock In {\em Proceedings of the 2008 USENIX Annual Technical Conference},
  ATC'08, pages 281--286, 2008.

\bibitem{Attariyan2010}
M.~Attariyan and J.~Flinn.
\newblock Automating configuration troubleshooting with dynamic information
  flow analysis.
\newblock In {\em Proceedings of the 9th USENIX Conference on Operating Systems
  Design and Implementation}, OSDI'10, pages 1--11, 2010.

\bibitem{BauerSACMAT2008}
L.~Bauer, S.~Garriss, and M.~K. Reiter.
\newblock Detecting and resolving policy misconfigurations in access-control
  systems.
\newblock In {\em Proceedings of the 13th ACM Symposium on Access Control
  Models and Technologies}, SACMAT '08, pages 185--194, 2008.

\bibitem{BeckettSIGCOMM2016}
R.~Beckett, R.~Mahajan, T.~Millstein, J.~Padhye, and D.~Walker.
\newblock Don't mind the gap: Bridging network-wide objectives and device-level
  configurations.
\newblock In {\em Proceedings of the 2016 ACM SIGCOMM Conference}, SIGCOMM '16,
  pages 328--341, Florianopolis, Brazil, 2016.

\bibitem{CadarKLEE2008}
C.~Cadar, D.~Dunbar, and D.~Engler.
\newblock {KLEE}: Unassisted and automatic generation of high-coverage tests
  for complex systems programs.
\newblock In {\em Proceedings of the 8th USENIX Conference on Operating Systems
  Design and Implementation}, OSDI'08, pages 209--224, San Diego, California,
  2008.

\bibitem{ChenCONEXT2010}
X.~Chen, Y.~Mao, Z.~M. Mao, and J.~Van~der Merwe.
\newblock Declarative configuration management for complex and dynamic
  networks.
\newblock In {\em Proceedings of the 6th International Conference}, Co-NEXT
  '10, pages 6:1--6:12, 2010.

\bibitem{ChipounovASPLOS2011}
V.~Chipounov, V.~Kuznetsov, and G.~Candea.
\newblock S2e: A platform for in-vivo multi-path analysis of software systems.
\newblock In {\em Proceedings of the Sixteenth International Conference on
  Architectural Support for Programming Languages and Operating Systems},
  ASPLOS XVI, pages 265--278, Newport Beach, California, USA, 2011.

\bibitem{DasSecurity2010}
T.~Das, R.~Bhagwan, and P.~Naldurg.
\newblock Baaz: A system for detecting access control misconfigurations.
\newblock In {\em Proceedings of the 19th USENIX Conference on Security},
  USENIX Security'10, pages 11--11, 2010.

\bibitem{DeTrevilleHOTOS2005}
J.~DeTreville.
\newblock Making system configuration more declarative.
\newblock In {\em Proceedings of the 10th Conference on Hot Topics in Operating
  Systems}, HOTOS'05, pages 11--11, 2005.

\bibitem{EnckATC2007}
W.~Enck, P.~McDaniel, S.~Sen, P.~Sebos, S.~Spoerel, A.~Greenberg, S.~Rao, and
  W.~Aiello.
\newblock Configuration management at massive scale: System design and
  experience.
\newblock In {\em Proceedings of the 2007 USENIX Annual Technical Conference},
  ATC'07, pages 6:1--6:14, 2007.

\bibitem{FeamsterNSDI2005}
N.~Feamster and H.~Balakrishnan.
\newblock Detecting {BGP} configuration faults with static analysis.
\newblock In {\em Proceedings of the 2nd Conference on Symposium on Networked
  Systems Design \& Implementation}, NSDI'05, pages 43--56, 2005.

\bibitem{twilio_july_2013}
Google.
\newblock Twilio billing incident post-mortem: Breakdown, analysis and root
  cause.
\newblock
  \url{https://www.twilio.com/blog/2013/07/billing-incident-post-mortem-breakdown-analysis-and-root-cause.html}.

\bibitem{Gray1986}
J.~Gray.
\newblock Why do computers stop and what can be done about it?
\newblock In {\em Proc. Symposium on Reliability in Distributed Software and
  Database Systems}, pages 3--12, 1986.

\bibitem{Herodotou11starfish}
H.~Herodotou, H.~Lim, G.~Luo, N.~Borisov, L.~Dong, F.~B. Cetin, and S.~Babu.
\newblock Starfish: A self-tuning system for big data analytics.
\newblock In {\em In CIDR}, pages 261--272, 2011.

\bibitem{HuangEuroSys2015}
P.~Huang, W.~J. Bolosky, A.~Singh, and Y.~Zhou.
\newblock {ConfValley}: A systematic configuration validation framework for
  cloud services.
\newblock In {\em Proceedings of the Tenth European Conference on Computer
  Systems}, EuroSys '15, pages 19:1--19:16, Bordeaux, France, 2015.

\bibitem{PerfContractsNSDI2019}
R.~Iyer, L.~Pedrosa, A.~Zaostrovnykh, S.~Pirelli, K.~Argyraki, and G.~Candea.
\newblock Performance contracts for software network functions.
\newblock In {\em Proceedings of the 16th USENIX Conference on Networked
  Systems Design and Implementation}, NSDI’19, page 517–530, Boston, MA,
  USA, 2019.

\bibitem{KellerDSN2008}
L.~Keller, P.~Upadhyaya, and G.~Candea.
\newblock {ConfErr}: A tool for assessing resilience to human configuration
  errors.
\newblock In {\em Proceedings of the 38th International Conference on
  Dependable Systems and Networks}, DSN'08, pages 157--166, 2008.

\bibitem{KingCACM1976}
J.~C. King.
\newblock Symbolic execution and program testing.
\newblock {\em Commun. ACM}, 19(7):385--394, July 1976.

\bibitem{Kushman2010OSDI}
N.~Kushman and D.~Katabi.
\newblock Enabling configuration-independent automation by non-expert users.
\newblock In {\em Proceedings of the 9th USENIX Conference on Operating Systems
  Design and Implementation}, OSDI'10, pages 1--10, 2010.

\bibitem{Lattner2004CGO}
C.~Lattner and V.~Adve.
\newblock {LLVM}: A compilation framework for lifelong program analysis \&
  transformation.
\newblock In {\em Proceedings of the 2004 International Symposium on Code
  Generation and Optimization}, CGO '04, pages 75--, Palo Alto, California,
  2004.

\bibitem{LearnConfEuroSys2020}
C.~Li, S.~Wang, H.~Hoffmann, and S.~Lu.
\newblock Statically inferring performance properties of software
  configurations.
\newblock In {\em Proceedings of the Fifteenth European Conference on Computer
  Systems}, EuroSys '20, Heraklion, Greece, 2020.

\bibitem{LooSIGCOMM2005}
B.~T. Loo, J.~M. Hellerstein, I.~Stoica, and R.~Ramakrishnan.
\newblock Declarative routing: Extensible routing with declarative queries.
\newblock In {\em Proceedings of the 2005 Conference on Applications,
  Technologies, Architectures, and Protocols for Computer Communications},
  SIGCOMM '05, pages 289--300, 2005.

\bibitem{Oppenheimer2003}
D.~Oppenheimer, A.~Ganapathi, and D.~A. Patterson.
\newblock Why do {Internet} services fail, and what can be done about it?
\newblock In {\em Proceedings of the 4th Conference on USENIX Symposium on
  Internet Technologies and Systems (USITS)}, Seattle, WA, Mar. 2003.

\bibitem{OsogamiSIGMETRIC2006}
T.~Osogami and T.~Itoko.
\newblock Finding probably better system configurations quickly.
\newblock In {\em Proceedings of the Joint International Conference on
  Measurement and Modeling of Computer Systems}, SIGMETRICS '06/Performance
  '06, pages 264--275, Saint Malo, France, 2006.

\bibitem{RabkinIEEE2013}
A.~Rabkin and R.~Katz.
\newblock How {Hadoop} clusters break.
\newblock {\em IEEE Softw.}, 30(4):88--94, July 2013.

\bibitem{UCSE2015Sec}
D.~A. Ramos and D.~Engler.
\newblock Under-constrained symbolic execution: Correctness checking for real
  code.
\newblock In {\em Proceedings of the 24th USENIX Conference on Security
  Symposium}, SEC’15, page 49–64, Washington, D.C., 2015.

\bibitem{schpbach2011a}
A.~Schüpbach, A.~Baumann, T.~Roscoe, and S.~Peter.
\newblock A declarative language approach to device configuration.
\newblock In {\em Proceedings of the 6th International Conference on
  Architectural Support for Programming Languages and Operating Systems},
  ASPLOS'11. ACM, March 2011.

\bibitem{Su2007SOSP}
Y.-Y. Su, M.~Attariyan, and J.~Flinn.
\newblock {AutoBash}: Improving configuration management with operating system
  causality analysis.
\newblock In {\em Proceedings of Twenty-first ACM SIGOPS Symposium on Operating
  Systems Principles}, SOSP '07, pages 237--250, 2007.

\bibitem{TangSOSP2015}
C.~Tang, T.~Kooburat, P.~Venkatachalam, A.~Chander, Z.~Wen, A.~Narayanan,
  P.~Dowell, and R.~Karl.
\newblock Holistic configuration management at facebook.
\newblock In {\em Proceedings of the 25th Symposium on Operating Systems
  Principles}, SOSP '15, pages 328--343, Monterey, California, 2015.

\bibitem{Wang2004OSDI}
H.~J. Wang, J.~C. Platt, Y.~Chen, R.~Zhang, and Y.-M. Wang.
\newblock Automatic misconfiguration troubleshooting with {PeerPressure}.
\newblock In {\em Proceedings of the 6th Conference on Symposium on Opearting
  Systems Design \& Implementation}, OSDI'04, pages 17--17, 2004.

\bibitem{SmartConf2018}
S.~Wang, C.~Li, H.~Hoffmann, S.~Lu, W.~Sentosa, and A.~I. Kistijantoro.
\newblock Understanding and auto-adjusting performance-sensitive
  configurations.
\newblock In {\em Proceedings of the Twenty-Third International Conference on
  Architectural Support for Programming Languages and Operating Systems},
  ASPLOS ’18, page 154–168, Williamsburg, VA, USA, 2018.

\bibitem{Wang2003LISA}
Y.-M. Wang, C.~Verbowski, J.~Dunagan, Y.~Chen, H.~J. Wang, C.~Yuan, and
  Z.~Zhang.
\newblock Strider: A black-box, state-based approach to change and
  configuration management and support.
\newblock In {\em Proceedings of the 17th USENIX Conference on System
  Administration}, LISA '03, pages 159--172, 2003.

\bibitem{RongATC203261}
X.~Wei, S.~Shen, R.~Chen, and H.~Chen.
\newblock Replication-driven live reconfiguration for fast distributed
  transaction processing.
\newblock In {\em 2017 {USENIX} Annual Technical Conference ({USENIX} {ATC}
  17)}, ATC 17, pages 335--347. {USENIX} Association, July 2017.

\bibitem{Whitaker2004OSDI}
A.~Whitaker, R.~S. Cox, and S.~D. Gribble.
\newblock Configuration debugging as search: Finding the needle in the
  haystack.
\newblock In {\em Proceedings of the 6th Conference on Symposium on Opearting
  Systems Design \& Implementation}, OSDI'04, pages 6--6, 2004.

\bibitem{XiWWW2004}
B.~Xi, Z.~Liu, M.~Raghavachari, C.~H. Xia, and L.~Zhang.
\newblock A smart hill-climbing algorithm for application server configuration.
\newblock In {\em Proceedings of the 13th International Conference on World
  Wide Web}, WWW '04, pages 287--296, New York, NY, USA, 2004.

\bibitem{PCheckOSDI16}
T.~Xu, X.~Jin, P.~Huang, Y.~Zhou, S.~Lu, L.~Jin, and S.~Pasupathy.
\newblock Early detection of configuration errors to reduce failure damage.
\newblock In {\em Proceedings of the The 12th USENIX Symposium on Operating
  Systems Design and Implementation}, OSDI '16, November 2016.

\bibitem{Xu2013SOSP}
T.~Xu, J.~Zhang, P.~Huang, J.~Zheng, T.~Sheng, D.~Yuan, Y.~Zhou, and
  S.~Pasupathy.
\newblock Do not blame users for misconfigurations.
\newblock In {\em Proceedings of the Twenty-Fourth ACM Symposium on Operating
  Systems Principles}, SOSP '13, pages 244--259, 2013.

\bibitem{Csmith2011PLDI}
X.~Yang, Y.~Chen, E.~Eide, and J.~Regehr.
\newblock Finding and understanding bugs in {C} compilers.
\newblock In {\em Proceedings of the 32nd ACM SIGPLAN Conference on Programming
  Language Design and Implementation}, PLDI ’11, page 283–294, San Jose,
  California, USA, 2011.

\bibitem{YeSIGMETRICS2003}
T.~Ye and S.~Kalyanaraman.
\newblock A recursive random search algorithm for large-scale network parameter
  configuration.
\newblock In {\em Proceedings of the 2003 ACM SIGMETRICS International
  Conference on Measurement and Modeling of Computer Systems}, SIGMETRICS '03,
  pages 196--205, San Diego, CA, USA, 2003.

\bibitem{YinSOSP2011}
Z.~Yin, X.~Ma, J.~Zheng, Y.~Zhou, L.~N. Bairavasundaram, and S.~Pasupathy.
\newblock An empirical study on configuration errors in commercial and open
  source systems.
\newblock In {\em Proceedings of the Twenty-Third ACM Symposium on Operating
  Systems Principles}, SOSP '11, pages 159--172, 2011.

\bibitem{YuanATC2011}
D.~Yuan, Y.~Xie, R.~Panigrahy, J.~Yang, C.~Verbowski, and A.~Kumar.
\newblock Context-based online configuration-error detection.
\newblock In {\em Proceedings of the 2011 USENIX Conference on USENIX Annual
  Technical Conference}, ATC'11, pages 28--28, 2011.

\bibitem{CDBTune2019}
J.~Zhang, Y.~Liu, K.~Zhou, G.~Li, Z.~Xiao, B.~Cheng, J.~Xing, Y.~Wang,
  T.~Cheng, L.~Liu, and et~al.
\newblock An end-to-end automatic cloud database tuning system using deep
  reinforcement learning.
\newblock In {\em Proceedings of the 2019 International Conference on
  Management of Data}, SIGMOD ’19, page 415–432, Amsterdam, Netherlands,
  2019.

\bibitem{ZhangASPLOS2014}
J.~Zhang, L.~Renganarayana, X.~Zhang, N.~Ge, V.~Bala, T.~Xu, and Y.~Zhou.
\newblock {EnCore}: Exploiting system environment and correlation information
  for misconfiguration detection.
\newblock In {\em Proceedings of the 19th International Conference on
  Architectural Support for Programming Languages and Operating Systems},
  ASPLOS '14, pages 687--700, 2014.

\bibitem{BestConfig2017}
Y.~Zhu, J.~Liu, M.~Guo, Y.~Bao, W.~Ma, Z.~Liu, K.~Song, and Y.~Yang.
\newblock {BestConfig}: Tapping the performance potential of systems via
  automatic configuration tuning.
\newblock In {\em Proceedings of the 2017 Symposium on Cloud Computing}, SoCC
  ’17, page 338–350, Santa Clara, California, 2017.

\end{thebibliography}
